\documentstyle[11pt]{article}
\begin{document}

\newcommand{\newc}{\newcommand}
\newc{\be}{\begin{equation}}
\newc{\ee}{\end{equation}}
\newc{\bea}{\begin{eqnarray}}
\newc{\eea}{\end{eqnarray}}
\newc{\bean}{\begin{eqnarray*}}
\newc{\eean}{\end{eqnarray*}}
\newc{\eqn}[1]{(\ref{#1})}
\newc{\gsim}{\lower.7ex\hbox{$\;\stackrel{\textstyle>}{\sim}\;$}}
\newc{\lsim}{\lower.7ex\hbox{$\;\stackrel{\textstyle<}{\sim}\;$}}
\newc{\ra}{\rightarrow}
\newc{\alphas}{\alpha_s}
\newc{\tanb}{\tan\beta}
\newc{\Lsoft}{{\cal L}_{soft}}
\newc{\flr}{ f_{L,R} }
\newc{\slr}{\widetilde l_R}
\newc{\hinot}{\widetilde H^0_2}
\newc{\hinob}{\widetilde H^0_1}
\newc{\bino}{\widetilde B^0}
\newc{\wino}{{\widetilde W^0_3}}
\newc{\gluino}{{\widetilde g}}
\newc{\mgluino}{m_{\gluino}}
\newc{\msl}{m_{\widetilde l}}
\newc{\msq}{m_{\widetilde q}}
\newc{\msf}{m_{\widetilde f}}
\newc{\mz}{m_Z}
\newc{\mw}{m_W}
\newc{\hl}{h}	\newc{\hh}{H}
\newc{\ha}{A}	\newc{\hcpm}{H^\pm}
\newc{\mhl}{m_h}	\newc{\mhh}{m_H}
\newc{\mha}{m_A}	\newc{\mhcpm}{m_{\hcpm}}
\newc{\rhocrit}{\rho_{\rm crit}}
\newc{\abund}{\Omega h_0^2}
\newc{\abundchi}{\Omega_\chi h_0^2}
\newc{\mchi}{m_\chi}
\newc{\mneutone}{m_{\chi^0_1}}	\newc{\mneuttwo}{m_{\chi^0_2}}
\newc{\neuttwo}{\chi^0_2}
\newc{\mcharone}{m_{\charone}}	\newc{\charone}{\chi_1^\pm}
\newc{\mhalf}{m_{1/2}}
\newc{\mzero}{m_0}
\newc{\muzero}{\mu_0}
\newc{\sgnmu}{{\rm sgn}\,\muzero}
\newc{\azero}{A_0}
\newc{\bzero}{B_0}
\newc{\mt}{m_t}
\newc{\mb}{m_b}
\newc{\mtau}{m_\tau}
\newc{\htop}{h_t}
\newc{\hbot}{h_b}
\newc{\htau}{h_\tau}
\newc{\mtpole}{m_t^{\rm pole}}
\newc{\mbpole}{m_b^{\rm pole}}
\newc{\mqpole}{m_q^{\rm pole}}
\newc{\mlpole}{m_l^{\rm pole}}
\newc{\mgut}{M_X}
\newc{\alphax}{\alpha_X}
\newc{\msusy}{M_{\rm SUSY}}
\newc{\msusyeff}{M^{eff}_{\rm SUSY}}
\newc{\ie}{{\it i.e.}}
\newc{\etal}{{\it et al.}}
\newc{\eg}{{\it e.g.}}
\newc{\etc}{{\it etc.}}
\newc{\superhu}{\hat H_u}	\newc{\superhd}{\hat H_d}
\newc{\superl}{\hat L}		\newc{\superr}{\hat R}
\newc{\superec}{\hat {e}^c}
\newc{\superq}{\hat Q}
\newc{\superu}{\hat U}
\newc{\superd}{\hat D}
\newc{\superuc}{\hat {u}^c}
\newc{\superdc}{\hat {d}^c}
\newc{\tildeQ}{\widetilde Q}
\newc{\tildeU}{\widetilde U}
\newc{\tildeD}{\widetilde D}
\newc{\tildeL}{\widetilde L}
\newc{\tildeuc}{\widetilde {u}^c}
\newc{\tildedc}{\widetilde {d}^c}
\newc{\tildeec}{\widetilde {e}^c}
\newc{\tildenu}{\widetilde \nu}
\newc{\MS}{{\rm\overline{MS}}}
\newc{\DR}{{\rm\overline{DR}}}
\newc{\stopq}{{\widetilde t}}		\newc{\stp}{\stopq}
\newc{\stopl}{\widetilde t_L}
\newc{\stopr}{\widetilde t_R}
\newc{\stopone}{\widetilde t_1}
\newc{\stoptwo}{\widetilde t_2}

\newc{\mstopq}{m_{\tilde t}}
\newc{\mstopl}{m_{\tilde t_L}}
\newc{\mstopr}{m_{\tilde t_R}}
\newc{\mstopone}{m_{\tilde t_1}}
\newc{\mstoptwo}{m_{\tilde t_2}}
\newc{\stau}{{\widetilde \tau}}
\newc{\staul}{\widetilde \tau_L}
\newc{\staur}{\widetilde \tau_R}
\newc{\stauone}{\widetilde \tau_1}
\newc{\stautwo}{\widetilde \tau_2}

\newc{\mstau}{m_{\tilde \tau}}
\newc{\mstaul}{m_{\tilde \tau_L}}
\newc{\mstaur}{m_{\tilde \tau_R}}
\newc{\mstauone}{m_{\tilde \tau_1}}
\newc{\mstautwo}{m_{\tilde \tau_2}}
\newc{\onehalf}{\frac{1}{2}}
\newc{\vhiggs}{V_{\rm Higgs}}
\newc{\delV}{\Delta V}
\newc{\sinsqthw}{\sin^2\theta_{\rm w}}
\newc{\ai}{\alpha_1}
\newc{\aii}{\alpha_2}
\newc{\ev}{{\rm\,eV}}
\newc{\mev}{{\rm\,MeV}}
\newc{\gev}{{\rm\,GeV}}
\newc{\tev}{{\rm\,TeV}}
\newc{\agut}{\alpha_X}
\newc{\Uy}{\mbox{U(1)}_{\rm Y}}
\newc{\Uem}{\mbox{U(1)}_{\rm EM}}
\newc{\vev}{\hbox{\it v.e.v.}}
\newc{\vone}{v_d}	\newc{\vtwo}{v_u}
\newc{\sbot}{{\tilde{b}}}
\newc{\slepton}{{\tilde{l}}}
\newc{\supq}{{\tilde{u}}}
\newc{\sdown}{{\tilde{d}}}
\newc{\selectron}{{\tilde{e}}}
\newc{\sneutrino}{{\tilde{\nu}}}
\newc{\squark}{{\tilde{q}}}
\newc{\higgsino}{{\tilde{H}}}
\newc{\bsg}{b\to s\gamma}
\newc{\mh}{\mhl}
\newc{\mx}{\mgut}
\newc{\eL}{\tilde{e}_L}
\newc{\eR}{\tilde{e}_R}

\hyphenation{pa-ram-e-trize pa-ram-e-tri-za-tion}
\hyphenation{Mad-i-son pre-prints}

\def\NPB#1#2#3{Nucl. Phys. B {\bf#1} (19#2) #3}
\def\PLB#1#2#3{Phys. Lett. B {\bf#1} (19#2) #3}
\def\PLBold#1#2#3{Phys. Lett. {\bf#1B} (19#2) #3}
\def\PRD#1#2#3{Phys. Rev. {\bf D#1} (19#2) #3}
\def\PRL#1#2#3{Phys. Rev. Lett. {\bf#1} (19#2) #3}
\def\PRT#1#2#3{Phys. Rep. {\bf#1} (19#2) #3}
\def\ARAA#1#2#3{Ann. Rev. Astron. Astrophys. {\bf#1} (19#2) #3}
\def\ARNP#1#2#3{Ann. Rev. Nucl. Part. Sci. {\bf#1} (19#2) #3}
\def\MODA#1#2#3{Mod. Phys. Lett. {\bf A#1} (19#2) #3}
\def\ZPC#1#2#3{Zeit. f\"ur Physik {\bf C#1} (19#2) #3}
\def\APJ#1#2#3{Ap. J. {\bf#1} (19#2) #3}

\begin{titlepage}

\begin{flushright}
{\large
UM-TH-93-24\\
October 1993\\
}
\end{flushright}
\vskip 2cm
\begin{center}
{\Large\bf STUDY OF CONSTRAINED MINIMAL SUPERSYMMETRY
}
\vskip 1cm
{\Large
G.L. Kane\footnote{E-mail: {\tt gkane@umich.edu}},
Chris Kolda\footnote{E-mail: {\tt kolda@umich.edu}},
Leszek Roszkowski\footnote{E-mail: {\tt leszekR@umich.edu}},
and
James D. Wells\footnote{E-mail: {\tt jwells@umich.edu}}
\\}
\vskip 2pt
{\large\it Randall Physics Laboratory, University of Michigan,\\ Ann Arbor,
MI 48190, USA}\\
\end{center}
\vskip .5cm
\begin{abstract}
Taking seriously the phenomenological indications for supersymmetry,
we have made a detailed study of unified minimal SUSY, including many effects
at the few percent level in a consistent fashion. We report here a general
analysis of what can be studied without choosing a particular gauge group at
the unification scale. Firstly, we find that the encouraging SUSY
unification results of recent years do survive the challenge of a more
complete and accurate analysis. Taking into account effects at the
5-10\% level leads to several improvements of previous results, and allows us
to sharpen our predictions for SUSY in the light of unification.
We perform a thorough study of the parameter space and look for patterns to
indicate SUSY predictions, so that
they do not depend on arbitrary choices of some parameters or
untested assumptions. Our results can be viewed as a fully constrained
Minimal SUSY Standard Model (CMSSM). The resulting model forms a well-defined
basis for comparing the physics potential of different facilities. Very little
of the acceptable parameter space has been excluded by LEP or FNAL so far, but
a significant fraction can be covered when these accelerators are upgraded.
A number of initial applications to the understanding of the values of
$\mhl$ and $\mt$, the SUSY spectrum, detectability of SUSY
at LEP~II or FNAL, BR($b\rightarrow s\gamma$),
$\Gamma(Z\ra b\bar b)$, dark matter,
\etc, are included in a separate
section that might be of more interest to some readers than the technical
aspects of model-building.
We formulate an approach to extracting SUSY parameters from data when
superpartners are detected. For small $\tanb$ or large $\mt$ both
$\mhalf$ and $\mzero$ are entirely bounded from above at $O(1\tev)$ without
having to use a fine-tuning constraint.

\end{abstract}
\end{titlepage}
\setcounter{footnote}{0}
\setcounter{page}{2}
\setcounter{section}{0}
\setcounter{subsection}{0}
\setcounter{subsubsection}{0}
\tableofcontents
\newpage


\section{Introduction}\label{intro:sec}

There has recently been a significant amount of activity in the field
of supersymmetric grand unification (SUSY GUT) and its possible
implications for the existence of low-energy SUSY and for future SUSY searches.
This renewed interest was
primarily caused by the observation~\cite{langackerluo} that LEP measurements
of the
gauge coupling constants seem to imply their (grand) unification
in a supersymmetric theory with superpartners near the weak scale, reinforced
by the
awareness that several phenomenological outcomes were consistent with
SUSY~\cite{susyrev}
although they need not have been. It was
shown~\cite{langackerluo,ekn,amaldi,anselmo} that the couplings merge at the
GUT energy scale $\mgut$
even in the simplest supersymmetric extension of the Standard
Model, the so-called MSSM, while they badly fail to do so in the
Standard Model (SM) alone.
This remarkable fact has been interpreted by many as a strong hint
for a SUSY GUT, especially since its main arch-rivals for expected
physics beyond the SM, the composite and technicolor approaches, seem
now even more disfavored by the precise measurements at LEP~\cite{deadend}.
It has been argued that the unification of gauge couplings
is also possible in some non-SUSY models~\cite{nonsusyguts}. These models are,
however,
exuberantly complicated and lack other virtues. Besides, one
should not forget that ordinary GUTs suffer from the hierarchy
and naturalness problems which SUSY automatically cures.

Certainly SUSY gauge coupling unification does not constitute
a proof
of SUSY, nor can it serve as a substitute for the direct discovery
of a SUSY particle. On the other hand, it is clearly very encouraging and
should not be ignored.
In fact, initial simplified studies~\cite{amaldi} claimed that it should be
possible to put stringent limits on $\mgut$ and the GUT value of the
gauge coupling $\alphax$, as well as on the typical scale of supersymmetry
breaking.
Subsequently it was realized~\cite{ekn,anselmo,rr,bh} that
additional effects, both around the electroweak and the GUT scale,
may introduce significant modifications to the early results without,
however, destroying supersymmetric unification.
Several authors thus focussed on increasingly refined studies
of the subtleties of gauge coupling
unification~\cite{ekn,anselmo,rr,lp1,hagiwara,bbo,meshkov,florida1}.
Some~\cite{ramondmbot,bbop,cpw,lp2} also considered the
unification of the bottom and tau masses, which, in addition to the
prediction of the correct value of $\sinsqthw$, was regarded
as a success of the early GUTs.
In some of these studies it was argued that the $\mb-\mtau$ unification
almost invariably implies a very heavy top quark. Many studies mentioned
above typically did not address other important issues of the MSSM.
(Some, for example, did not require
correct electroweak gauge symmetry breaking.)
Finally, some studies have adopted a more comprehensive
approach~\cite
{rr,angut,roberts,klnpy1,klnpy2,lnz:dm,dn,deboer,florida2,bbonew}.
The goal is to generate, simultaneously with gauge coupling unification,
realistic mass spectra of the Higgs and SUSY particles.
This is usually done in the framework of the MSSM coupled to
the minimal supergravity model which relates
many unknown quantities of the MSSM in terms of a few basic
parameters at the GUT scale.
Next, various experimental and cosmological limits can be applied
to the resulting couplings and mass spectra. One can then examine whether all
the
constraints are consistent with each other and whether the SUSY
partners have masses in the region of low-energy ($\lsim1\tev$) SUSY.
This is the way we study the MSSM in this work. Similar approaches have
been studied in the programs of Arnowitt and Nath in
Refs.~\cite{angut,andm,anscaling} and of
Lopez, Nanopoulos, \etal, in
Refs~\cite{klnpy1,klnpy2,lnz:dm,ln:proton,sinalpha}
among others.

We want to stress that only such a comprehensive study can be regarded
as relatively self-consistent. Considering gauge coupling unification alone
neglects the contribution (at 2-loops) from the Yukawa couplings.
It usually also assumes grossly oversimplified supersymmetric mass
spectra. More importantly, if one wants to include also the running of the
Yukawa
couplings, one is faced with the problem of whether or not one can at the same
time
generate electroweak symmetry breaking~\cite{gsymbreak}, where the magnitude of
the top
Yukawa coupling is of crucial importance. Furthermore, in general one
must take into account the running of {\em all} the Yukawa coupling of
the third generation, as we do in the present study.
In order to impose electroweak symmetry breaking properly
one needs to run not only the Higgs mass parameters but in fact all the
relevant SUSY
parameters which will be specified below. Deriving spectra that are compatible
with
the gauge coupling unification and electroweak symmetry breaking can only be
achieved if the whole set of relevant parameters is simultaneously evaluated.
Further, numerical effects from 2-loops, the full 1-loop Higgs effective
potential, \etc, often significantly affect the results.

Only after implementing this comprehensive
approach are we able to reject the ranges of parameters that are either
unphysical or experimentally excluded, while maintaining
consistency with gauge coupling unification.
Many of the detailed effects we include have important consequences. For
example, the two typical solutions presented as a result of such an
analysis in Ref.~\cite{rr} are no longer acceptable when the more complete
analysis is done.

Once we derive a self-consistent SUSY spectrum that follows from grand
unification,
we can compare it with the present experimental limits. Furthermore,
we can study its implications for cosmology and derive additional bounds.
Finally, we may establish what ranges
(and properties) of the parameter space are compatible with all
limits. Such ranges should then be focussed on in planning for
experimental searches
as the most ``natural'' --- those expected in the Constrained Minimal
Supersymmetric Standard Model (CMSSM).

It should be emphasized that one reason it is worthwhile doing extensive work
constructing SUSY models and analyzing their implications even though the full
theory is not known, and the origin of SUSY breaking
is not understood, is that the form of the Lagrangian at the GUT energy
scale ($\sim 10^{16}\gev$) is very general and quite insensitive to our
ignorance.
The kinetic energy terms are not completely unique, but corrections are likely
to be of order $v_{\rm GUT}/m_{\rm Pl}$ and thus
small~\cite{hkq}. Apart from these, given the $R$-parity conservation that we
think is motivated by the stability of the proton and by cold dark matter, the
superpotential we write is general, and so is the form of the soft
terms~\cite{grisaru}. Whether one arrives at the Lagrangian from supergravity
or string
theory, it has the same form~\cite{stringsoft} so long as quadratic divergences
that would mix the high and low scales  (\ie, terms which are
not soft) are excluded. Thus anything one can learn about the Lagrangian by
imposing physics constraints will be of general validity. Until superpartners
are detected the information we have will not be sufficiently extensive to
determine all parameters in it separately, of course, so one will have to make
various simplifying assumptions. These assumptions can be tested in many ways
as soon as superpartner masses and branching ratios
are available.

Some aspects of SUSY GUTs, most notably the GUT-scale
corrections~\cite{hagiwara,lp2,angut,hrs}
to the running of the gauge and Yukawa couplings and proton
decay~\cite{angut,ln:proton,protonalts}, can only be considered once a
specific GUT model is selected.
While we have no objections to most GUT gauge groups, for several reasons we
would rather proceed  by first learning what we can say
without specifying a gauge group, and then by making a comparative study of GUT
gauge groups. One reason is that there may be no unification group at
all~\cite{ibross}.
In fact in many string models the SM gauge group (perhaps enlarged by one or
two
$U(1)$'s) is obtainable directly from
strings in which case one has gauge coupling unification without an
underlying gauge group unification. Also we are
concerned about $SU(5)$ as a unification gauge group
because we think it would be an astonishing accident if SUSY was otherwise
successful and also provided just the amount and kind of cold
dark matter needed by cosmology, but either Nature did not use this dark matter
or did not have it occur naturally~\cite{martin}\ in the structure of the
theory (as would have to be the case with $SU(5)$ because $R$-parity
conservation has to be imposed by hand there). Further, while some
groups~\cite{angut,ln:proton} have shown that the proton decay
constraint can be very important, others~\cite{protonalts} have argued that
the situation is not unambiguous. Thus we feel that it is important
to maintain the distinction between the MSSM and the particular low-energy
model one derives by choosing a specific GUT model.
We will assume throughout this analysis only that the gauge couplings unify
with $\sinsqthw(\mx)=3/8$ and that the theory remains $SU(3)\times SU(2)\times
U(1)_Y$ symmetric up to that unification scale.
We are extending our approach to include a comparative study  of implications
of unification gauge groups (or no simple unification), and will report on this
in a future publication.

In Section~\ref{basics:sec} we briefly remind the reader of the basic
assumptions underlying the MSSM. In Section~\ref{thresholds:sec} we
take an initial approach to the issue of gauge coupling unification
and focus in particular on the effect of light mass thresholds.
In Section~\ref{mbtau:sec} we digress on the issue of $\mb-\mtau$ unification
and discuss to what extent it requires a very heavy top quark. The dynamical
radiative electroweak symmetry breaking and the resulting
constraints are treated in Section~\ref{ewsb:sec}. In Section~\ref{sugra:sec}
we briefly list supergravity-induced relations
between the parameters of the model, and in the next Section we use them to
specify the list of independent parameters that
we choose to perform our numerical studies. Also in this Section we describe
the technical aspects of the procedure used in this analysis. In
Section~\ref{constraints:sec} we discuss several experimental and cosmological
limits which we use in Section~\ref{results:sec} to constrain the remaining
parameter
space. In Section~\ref{results:sec} we also survey a number of
results of our analysis  concerning the resulting patterns of SUSY
spectra. From the phenomenological point of view we arrive at a COnstrained
Minimal PArameterS Space (COMPASS) such
that every choice of constrained parameters is guaranteed to have gauge
coupling unification, electroweak symmetry breaking, and all experimental
constraints and cosmological constraints satisfied. COMPASS
will be our guide to what predictions could really occur and are not excluded
by any known constraint.

COMPASS still does not uniquely determine each parameter ($\mhalf$, $\mzero$,
$\mt$,
$\tanb$, $\azero$, $\sgnmu$ as defined in Section~\ref{procedure:sec}).
They can take a range of (highly correlated) values, though remarkably it
typically
implies mass spectra within the 1\tev\ mass range.
We want to avoid further assumptions about the parameters because no further
theory or data is available to guide us, so
we explore the general implications resulting from COMPASS
by varying all relevant parameters over wide ranges of values. In future work
we
will explore in detail predictions for hadron (FNAL, LHC) and electron (LEP,
LEP~II, NLC) colliders, including to what extent  SUSY is detectable  at LEP~II
and FNAL (with upgrades); in Section~\ref{implications:sec} we give a first
survey. We also study such
issues as what gives the dominant contributions to $\mhl$ and $\mt$, the
spectrum of superpartners and predictions of SUSY for the cosmological
abundance of the lightest supersymmetric particle (LSP),
BR$(b\rightarrow s\gamma)$ and $\Gamma(Z\ra b\bar b)$.
In addition we briefly illustrate a new approach to extracting SUSY parameters
from data. Solving the equations giving the parameters of the Lagrangian in
terms of experimental observables can be difficult
and misleading if approximations are introduced, but with our CMSSM the basic
parameters can be easily extracted.
Section~\ref{implications:sec} can be
read independently of the rest of the paper, and those more interested
in the phenomenological implications rather than the technical aspects  of
model-building may prefer to do so. Although this paper is long, we
think it is very important to present a single treatment that generates
solutions of the CMSSM consistent with all theoretical and experimental
constraints, and
examines their consequences and predictions.


\section{Formalism}\label{formal:sec}

\subsection{Basic Assumptions}\label{basics:sec}

Several features make the Minimal Supersymmetric Standard Model (MSSM)
a particularly interesting extension of the Standard Model.
The model is based on the same gauge group as the SM, and its particle
content is the minimal one required to implement supersymmetry
in a consistent way. It is described by the $R$-parity conserving
superpotential
\be
W= h^U_{ij} \superq_i\superhu\superuc_j
           + h^D_{ij} \superq_i\superhd\superdc_j
	   + h^E_{ij} \superl_i\superhd\superec_j
	   + \mu \superhd\superhu.
\label{spotential:eq}
\ee
Here $\superq$, $\superl$ represent the quark and lepton $SU(2)$ doublet
superfields, $\superuc$, $\superdc$, $\superec$ the corresponding $SU(2)$
singlets, and $\superhu$, $\superhd$ the Higgs superfields whose scalar
components give mass to up- and down-type quarks/leptons respectively.
Generational indices have been shown explicitly, but group indices have been
dropped. In addition, one introduces all the allowed
soft supersymmetry-breaking terms. These are given by
\bea
-\Lsoft
& = & \left(A^U_{(ij)} h^U_{ij} \tildeQ_i H_u \tildeuc_j
    + A^D_{(ij)} h^D_{ij} \tildeQ_i H_d \tildedc_j
    + A^E_{(ij)} h^E_{ij} \tildeL_i H_d \tildeec_j + h.c. \right)  \nonumber \\
&   & \mbox{} + B\mu \left(H_d H_u + h.c. \right)
    + m_{H_d}^2\vert H_d\vert^2 + m_{H_u}^2\vert H_u\vert^2 \nonumber\\
&   & \mbox{} + m_{\tildeL}^2\vert \tildeL\vert^2 + m_{\tildeec}^2\vert
\tildeec\vert^2
     + m_{\tildeQ}^2\vert \tildeQ\vert^2 + m_{\tildeuc}^2\vert \tildeuc\vert^2
     + m_{\tildedc}^2\vert \tildedc\vert^2 \nonumber \\
&   & \mbox{} + \left(\onehalf M_1 \bar{\psi}_B\psi_B + \onehalf
	M_2\bar{\psi}^a_W
	\psi^a_W + \onehalf \mgluino\bar{\psi}^a_g\psi^a_g + h.c. \right).
\label{lsoft:eq}
\eea
Here the tilded fields are the scalar partners of the quark and lepton
fields, while the $\psi_i$ are the spin-$\onehalf$ partners of the
$i=U(1)_Y, SU(2)_L, SU(3)_c$ gauge bosons. The $A_{(ij)}$, $B$, and all other
new parameters in $\Lsoft$ are {\em a priori} unknown mass parameters.

The full Lagrangian consists of the kinetic and gauge terms (which are assumed
to
be minimal), the terms derived from the superpotential (the $F$-terms), and
$\Lsoft$.
It is important to understand that the
Lagrangian we study has the most general set of $R$-parity conserving
soft-breaking terms, that is,
terms that do not induce quadratic divergences and thereby preserve the
existence of two disparate mass scales. We require $R$-parity conservation
motivated not only by the lack of fast proton decay in nature, but also by
the natural success of the theory in predicting the existence of dark
matter. In Section~\ref{sugra:sec} we will add some other assumptions that
relate various soft-breaking terms; these assumptions are somewhat motivated
and can easily be removed for further study if theoretical or
phenomenological opportunities exist.

The model as defined by Eqs.~(\ref{spotential:eq}), (\ref{lsoft:eq})
is the simplest phenomenologically
viable supersymmetric extension of the SM. It is also general in the
sense of allowing the most general form~\cite{grisaru} of soft terms in
Eq.~(\ref{lsoft:eq}). On
the other hand, because of the large number of new unknown
parameters the model is not very predictive. A natural way of relating
them is to think of the MSSM as coming out of some
underlying GUT (or string) model.

One possible approach is to select at the start a specific GUT which
at low-energy would take the form of the MSSM (plus possibly a modified
neutrino sector which we neglect here). This can be done
with any GUT which can break into the SM gauge group, the minimal
$SU(5)$ being the simplest and most often studied choice. In this approach,
however, one must also consider the whole GUT-scale structure with a more
complicated Higgs sector. Guided by minimality, one often focuses on the
simplest Higgs sector of $SU(5)$. But that model cannot be regarded as
realistic or particularly attractive due
to the well-known problem of doublet-triplet splitting. Fixing this
new ``fine-tuning'' problem at the GUT scale requires significant modifications
of the model. In other words, at present we believe there is no commonly
accepted ``standard'' GUT model.

Another approach is to treat the MSSM as an effective model that
could arise from a large class of GUT models while not making any specific
choice. Instead, one can make various reasonable assumptions at
the GUT scale consistent with general properties of that class of GUT
models and next study ``corrections'' due to a specific GUT.
In this approach one therefore initially neglects all possible corrections due
to the superheavy states.
This is the approach that we will follow here. We will be adopting more
and more assumptions at the GUT scale, starting in the next section
from just gauge coupling unification and eventually considering
the MSSM in the framework of the minimal supergravity model.
While we won't choose
any specific GUT we will remark below about the importance of
some of the possible corrections at the GUT energy scale. We feel it is
important to distinguish what we can learn from this approach from the results
that would be obtained if we chose a specific unification gauge group.

\subsection{Light Threshold Corrections}\label{thresholds:sec}

We first address several issues that can be studied without necessarily
introducing further simplifications of the parameter space. We begin by
focussing on the running of the gauge couplings alone and in particular on
the important role played by the mass thresholds due to the Higgs and
supersymmetric particles.

In running the Renormalization Group Equations (RGEs) between the
weak and GUT scales
the coefficients of the RGEs change at each particle's mass threshold due to
the
decoupling of states at scales above their masses.
Initially, a simplified case was considered~\cite{amaldi} where
one assumed mass
degeneracy for all the sparticles (along with the second Higgs
doublet) at some scale
usually denoted $\msusy$. In that case one uses the $\beta$-functions
for the gauge couplings of the SM between $Q=\mz$ and $Q=\msusy$, and
those of the MSSM between $Q=\msusy$ and $Q=\mgut$.

However, the effects of a non-degenerate SUSY spectrum on the
gauge coupling $\beta$-functions provides a significant
correction to the naive solutions of the
RGEs~\cite{anselmo,rr,ekngutcorrs,hmy,haberh}.
The assumption that $\msusy$ could represent
some average sparticle mass for highly non-degenerate
spectra, such as one gets
in super-unified models, is in general incorrect and can lead to significant
errors on the order of 10\% or more in $\alphas(\mz)$, $\mx$, \etc\
Instead one must take into account the various sparticle
thresholds individually, changing the gauge coupling $\beta$-function
coefficients
for each sparticle as the energy scale crosses its (running) mass, \ie, when
it decouples from the RGEs.
Accounting for each particle's contribution to the gauge
$\beta$-function coefficients, one can write at 1-loop~\cite{anselmo,rr}:
\bea
\lefteqn{b_1^{\rm MSSM}=\frac{4}{3}N_g + \frac{1}{10}N_H^{\rm SM}
		+ \frac{2}{5}\theta_{\higgsino}+\frac{1}{10}\theta_{H_2}}
		  \nonumber \\
         & &\mbox{}+\frac{1}{5}\sum_i\left\lbrace\frac{1}{12}\left(
	        \theta_{\supq_{L_i}}
		+ \theta_{\sdown_{L_i}}\right)
		+ \frac{4}{3}\theta_{\supq_{R_i}}
		+ \frac{1}{3}\theta_{\sdown_{R_i}}
		+ \frac{1}{4}\left(\theta_{\selectron_{L_i}}
		+ \theta_{\sneutrino_{L_i}}\right)
		+ \theta_{\selectron_{R_i}}\right\rbrace \label{eq:b1} \\
\lefteqn{b_2^{\rm MSSM}=-\frac{22}{3} + \frac{4}{3}N_g +
                \frac{1}{6}N_H^{\rm SM}
		+ \frac{4}{3}\theta_{\widetilde W}
		+ \frac{2}{3}\theta_\higgsino + \frac{1}{6}\theta_{H_2}
		\nonumber }\\
         & &\mbox{}+\frac{1}{2}\sum_i\left\lbrace\theta_{\supq_{L_i}}
	        \theta_{\sdown_{L_i}}
		+ \frac{1}{3}\theta_{\selectron_{L_i}}
		  \theta_{\sneutrino_{L_i}}\right\rbrace \label{eq:b2} \\
\lefteqn{b_3^{\rm MSSM}=-11 + \frac{4}{3}N_g + 2\,\theta_\gluino
		+ \frac{1}{6}\sum_i\left\lbrace\theta_{\supq_{L_i}}
		+ \theta_{\sdown_{L_i}}
		+ \theta_{\supq_{R_i}} + \theta_{\sdown_{R_i}}\right\rbrace}
		\label{eq:b3}
\eea
where
\be
\frac{d\alpha_i}{d t}\equiv\frac{b_i}{2\pi}\alpha_i^2
+ \mbox{\rm 2-loops},
\mbox{~~~~~}t\equiv\log(Q/\mz),\mbox{~~~~~}\ai\equiv\frac{5}{3}\alpha_Y,
\mbox{~~~~~}\theta_x\equiv\theta(Q^2-m_x^2).
\ee
In the summations, $i=1,\ldots,N_g$ where $N_g=3$ is the number of fermion
generations,
and $N_H^{\rm SM}=1$ is the number of SM Higgs doublets. Here also
$\higgsino$ represents the (mass degenerate) higgsino fields,
$\widetilde{W}$ the partners of the $W$-bosons ($m_{\widetilde{W}}\equiv M_2$),
and $\gluino$ the
partner of the gluon, all taken to be mass eigenstates in this
approximation. $H_2$ is to be understood as the second Higgs doublet in the
approximation
where $H_1$ is the SM Higgs doublet containing the neutral $CP$-even Higgs
boson with
mass $\sim\mz$. $H_2$ is heavy with each component's mass equal to
that of the pseudoscalar Higgs. In this approximation the mixing of the
two Higgs doublets is suppressed by inverse powers of the heavy Higgs bosons'
masses and are therefore ignored as being of higher order and numerically
negligible~\cite{haberh}.
(The full 2-loop gauge
coupling $\beta$-functions for the SM and MSSM which we use in actual
calculations can be found in Refs.~\cite{susyRGEs} and~\cite{smRGEs}
respectively.
A discussion of 2-loop thresholds can be found in Section~\ref{running:sec}.)

The effect of multiple mass threshold effects on the running of the
gauge couplings has been extensively studied recently.
Notably, in a semi-analytic
approach developed by Langacker and Polonsky~\cite{lp1}
the effects of the thresholds on the
1-loop gauge $\beta$-functions were studied. They argued that
in the calculation of $\alphas(\mz)$
from $\sinsqthw$, $\alpha$, and the GUT-unification condition, the net
effect of all low-energy thresholds could
be expressed in terms of a
single scale $\msusyeff$ (called $A_{\rm SUSY}$ in Ref.~\cite{lp1}).
One can express this scale in terms of all the supersymmetric masses:
\bea
\msusyeff &=& m_{\widetilde H}
\left({{m_{H_2}}\over{m_{\widetilde H}}}\right)^{3\over19}
\left({{M_2}\over{m_{\widetilde H}}}\right)^{4\over19}
\left({{M_2}\over{m_{\widetilde g}}}\right)^{28\over19}
\left(m_{\widetilde u_L} m_{\widetilde c_L} m_{\widetilde
t_L}\right)^{-{15\over114}}\\
& &\times\left(m_{\widetilde d_L} m_{\widetilde s_L} m_{\widetilde
b_L}\right)^{1\over2}
\left(m_{\widetilde u_R} m_{\widetilde c_R} m_{\widetilde
t_R}\right)^{-{15\over57}}
\left(m_{\widetilde d_R} m_{\widetilde s_R} m_{\widetilde
b_R}\right)^{-{9\over57}}
\nonumber \\
& &\times\left(m_{\widetilde e_L} m_{\widetilde \mu_L} m_{\widetilde
\tau_L}\right)^{7\over38}
\left(m_{\widetilde e_R} m_{\widetilde \mu_R} m_{\widetilde
\tau_R}\right)^{-{2\over19}}
\left(m_{\widetilde \nu_{eL}} m_{\widetilde \nu_{\mu L}}
m_{\widetilde \nu_{\tau L}}\right)^{-{1\over38}}.\nonumber
\label{msusyeff:eq}
\eea

In the simplified case in which the spectra of squarks and sleptons are each
assumed
mass-degenerate, and taking only the contributions with leading exponents,
Eq.~(\ref{msusyeff:eq}) reduces
to a similar formula given in Ref.~\cite{cpw}.
By using a very crude parametrization in which $m_\higgsino\simeq|\mu|$, one
finds that
\be
\msusyeff \simeq |\mu|\left(\frac{\alpha_2(M_2)}{\alphas(m_\gluino)}\right)
^{\frac{28}{19}} \simeq |\mu|/5.
\ee
This strong dependence on $\mu$ is somewhat unexpected considering that
$\mu$ does not break supersymmetry.

The $\msusyeff$ formalism is useful in providing estimates of
the size of the various possible corrections to the running of the gauge
couplings. However, it is neither accurate nor practical in the more
comprehensive approach that we will adopt below in which the running
of gauge couplings is simultaneously considered with the running of
Yukawa couplings and mass parameters. Using
the SM RGEs between $Q=\mz$ and $Q=\msusyeff$ and the SUSY RGEs for
$Q>\msusyeff$ may accurately reproduce $\alphas(\mz)$, but it
will not provide the correct
value of $\agut$ or $\mgut$ (up to 50\% errors for the latter),
nor will it allow one to calculate
correctly the ratio $\mb/\mtau(\mgut)$. Furthermore, in this scheme
2-loop corrections to the 1-loop value of $\alphas(\mz)$ derived in this method
can be added only in an approximate fashion.
These 2-loop corrections are of the same order as the
1-loop threshold corrections, and in fact tend to increase $\alphas(\mz)$ by
$O(10\%)$ when included (see Table~\ref{loops:table}
in Section~\ref{running:sec}). Thus we will not use the technique of an
effective SUSY
scale except for purposes of comparison in Section~\ref{mbtau:sec}.

In the numerical analysis that we will present later, the effect of
the threshold corrections is automatically included separately for each
contributing particle, not with a single SUSY threshold. We will discuss
this, along with some other subtleties involved, in Section~\ref{running:sec}.

Finally, several authors have also emphasized the importance
of thresholds at the GUT
scale~\cite{bh,lp1,hagiwara,meshkov,lp2,angut,hrs}. In many models, such as
minimal $SU(5)$,
these corrections can be sizeable. In fact, they can be comparable to
the corrections coming from the non-degeneracy of
the SUSY spectrum at the low scale (see, \eg, Ref.~\cite{lp1}).
Consideration of such corrections can even be used to
achieve gauge coupling unification
in models where none seemed otherwise possible, such as
non-supersymmetric SO(10)~\cite{langackerluo,nonsusyguts}.
Models with non-minimal GUT sectors  often give rise to sizable corrections
that can alter low-energy predictions~\cite{langackerluo,nmgutsector}. However,
consideration
of these corrections
can only be made after {\it (i)} a GUT gauge group has been
chosen and {\it (ii)} the GUT Higgs sector and mass spectrum has been
decided upon. Because we wish to study
the super-unified MSSM in general, without reference
to a particular choice of GUT gauge
group or spectrum, we ignore all such corrections and
leave them for future studies of
various proposed unification schemes.

\section{Bottom-Tau Yukawa Unification} \label{mbtau:sec}

There has been much interest recently in the issue of Yukawa
coupling unification
within the framework of SUSY. In many GUT models, including minimal
SUSY--$SU(5)$, the down-type components of the
lepton and quark doublets reside in the same GUT multiplets and, assuming a
particularly simple Higgs sector, their Yukawa couplings are often equal at
the GUT scale. The experimentally determined ratio
$\mb/\mtau\simeq3$, which decreases roughly to one at the GUT scale, was
considered one of the early successes for GUTs. More recently however after
the precise LEP data on gauge couplings became available, it was shown that
the bottom-tau mass unification, while consistent with SUSY--$SU(5)$, was
inconsistent with the non-SUSY case~\cite{ramondmbot}.

Several groups~\cite{bbop,cpw,lp2}
have examined $b-\tau$ mass
unification more precisely in
minimal SUSY under the assumption of gauge coupling unification.
These studies have claimed that in order to achieve $b-\tau$ mass unification
one must have a top quark with mass very near to its IR pseudo-fixed point.
That is, to a good approximation $b-\tau$ mass unification implies~\cite{bbop}:
\be
\mtpole\simeq(200\gev) \sin\beta. \label{eq:mtfix}
\ee
For top quark masses favored by LEP
$(130\gev\lsim\mtpole\lsim 170\gev)$~\cite{lepreview}
under the assumption of a light Higgs, they find
that only the small regions $1\lsim\tan\beta\lsim2$ or $\tan\beta\simeq60$ are
consistent with $b-\tau$ mass unification. (Here $\mtpole$ refers to the
so-called pole mass of the top quark as opposed to the running or $\MS$
mass~\cite{pole}. For a clear discussion of this point see
Ref.~\cite{florida1}.
We will usually speak of running masses except where we specify otherwise.)

One is led to ask: if the top quark is
found to have a mass somewhere in the LEP-favored region, are we absolutely
forced to either very small or very large values for $\tan\beta$? In order to
answer this question one must consider how stable the stated claim is to
perturbations in the inputs of the analysis. Such questions have been briefly
considered in Refs.~\cite{bbop,cpw,lp2,bbo:bostontalk}. We find that the
effects of such
perturbations are often understated.

In considering how to make the MSSM consistent with a ``light'' top quark
(\ie\ one with mass well below that required by Eq.~(\ref{eq:mtfix})),
we find that there are several options for eluding the
heavy top or the extreme values of $\tan\beta$
without having to give up on $b-\tau$ mass unification completely. First and
foremost, it must be remembered that previous attempts to address this
issue have suffered from a common problem: they have
attempted to study $b-\tau$ mass unification while using only a single
threshold
approximation for the SUSY mass spectrum. That is, these analyses have
claimed that a non-degenerate spectrum of sparticles can be approximated
by a single effective scale. Though this is indeed possible for a study
of $\alphas(\mz)$ consistent with gauge coupling unification (see discussion
in Section~\ref{thresholds:sec}), no
single threshold approximation can possibly perform the same task for
$b-\tau$ mass unification, given the dependence of the Yukawa RGEs on
the gauge couplings, the presence of Yukawa couplings in the 2-loop RGEs, and
the
necessary lack of knowledge about the
scale of unification in such an approximation.

With this caveat in mind we now begin to explore the stability
of Eq.~(\ref{eq:mtfix}) to perturbations in the inputs to the analysis.
In this section alone we shall use the very same single
threshold approximation about which we have just warned
the reader. We do so because we are only interested in general numerical
studies that point to possible approaches to this question, and because
we have a consistent approach in the following sections
whose results do not depend on the single threshold approximation for
the gauge couplings.

The one scale that we use here should not be confused with the $\msusy^{eff}$
introduced earlier, for we will choose $\alphas(\mz)$ in this case
without regard to
the condition of gauge coupling unification. This new effective scale is in
fact nothing more than the naive SUSY scale used in the studies of
Ref.~\cite{bbop}
and in many early SUSY studies. In displaying our results, we will choose this
effective SUSY scale to be equal to $\mz$; once again the exact value is
unimportant for our general conclusions. We also have to
choose a value for the $b$-quark pole mass. In this Section, we will
take the range $4.7\leq\mbpole\leq 5.1\gev$, which is
the $3\sigma$ bound from the recent analysis of Ref.~\cite{ty}.

In addition to the innate error resulting from a single threshold
approximation, there remain other simple routes by which Eq.~(\ref{eq:mtfix})
can be modified. We find that by
{\it (i)} allowing corrections to the Yukawa unification, or {\it (ii)}
allowing the strong coupling constant to take on values near the lower end
of its experimental range, one can avoid the requirement of a heavy top.

The first of these routes requires one to consider corrections to the
requirement
that $\mb/\mtau=1$ at $\mgut$. This is because in the interesting regions
of the $\mt-\tan\beta$ plane, one finds that in general $\mb/m_\tau<1$ at
$\mgut$. Corrections could be induced through radiative corrections,
through effects from heavy state decoupling,
through non-renormalizable operators, or simply
by the scale of $b-\tau$ mass unification becoming displaced from the scale of
gauge coupling unification. Without choosing any particular source, such
corrections have been considered~\cite{bbop,lp2,bbo:bostontalk}.
But how large must these corrections
become in order to significantly alter the central claim of
Eq.~(\ref{eq:mtfix})?
In Fig.~\ref{mbone:fig}, we have shown the
regions consistent with $\mb/\mtau=1$ for bottom quark masses
in the range $4.7\leq\mbpole\leq 5.1\gev$ and $\alphas(\mz)=0.120$
(within the solid lines). We have also shown the region for
the same range of bottom masses, but now with corrections to Yukawa
unification of 10\% (dashed lines). This or similar plots are most often
shown in the literature as evidence for the stability of Eq.~(\ref{eq:mtfix}).

Although Fig.~\ref{mbone:fig} suggests that Eq.~(\ref{eq:mtfix}) is stable
to a 10\% correction
one might also wish to explore the effect of varying the strong
coupling constant on the Yukawa unification. Current measurements of
$\alphas(\mz)$ from a variety of sources indicates that
$0.110\lsim\alphas(\mz)\lsim
0.130$. Values of $\alphas(\mz)$ in the lower half of this range in combination
with a 10\% uncertainty in the GUT relation $\mb/\mtau=1$ significantly widen
the available parameter space in the $\mt-\tan\beta$ plane.
(Given the analyses of Ref.~\cite{kramer},
such low values for $\alphas(\mz)$
should be included in a careful consideration of these questions.)
Such an effect is shown in Fig.~\ref{mbtwo:fig} where we have taken
$\alphas(\mz)=0.112$. It must be emphasized that such a small value for
$\alphas(\mz)$
is inconsistent with the simplest SUSY-GUT unification
unless we require the scale of
SUSY masses to be $O(10\tev)$. In particular, we would need a very heavy
higgsino. Nonetheless, such a small $\alphas(\mz)$ could
come from other sources, such as heavy threshold effects or
non-renormalizable operators. Fig.~\ref{mbtwo:fig} clearly shows that for
$\mt\gsim 140\gev$, all values of $\tan\beta$ consistent with
perturbative unification become allowed. Essentially, combining two 10\%
effects has eliminated the constraint among $\mb$, $\mt$, and $\tanb$.

As we have tried to emphasize, the conclusions drawn by demanding
strict $b-\tau$ mass
unification can be quite strong, yet fairly small
effects due to unknowns in the
analysis can change the results considerably. Therefore we take the
following approach in the remainder of this paper. We will always take the
$\tau$-mass as given very precisely by experiment and use it to determine
$\mtau(\mgut)$. Though the experimental uncertainty to the central value of
$4.9\gev$ is larger, we will do the same for the bottom quark mass.
We will not demand exact $b-\tau$ mass unification. Because
we make no specific choice of GUT group or spectrum in this paper,
we have no mechanism
otherwise for escaping the constraints imposed by
$b-\tau$ mass unification. Yet we also understand that corrections that will
come from any eventual choice of GUT can, as we demonstrated above, allow a
larger region of parameter space to become available. When we do require exact
$b-\tau$ mass unification in our full analysis, we find agreement with
Eq.~(\ref{eq:mtfix}). Further, even when we do not require unification, all
solutions generated still preserve unification to about 20\%.

Because we wish this analysis to be general and to provide
insights over the entire range of perturbatively allowed values of
$\tan\beta$ in particular, we must do without
precise $b-\tau$ mass unification.
At the same time we still include everything that would otherwise follow from
imposing this unification because solutions we find in the regions of parameter
space consistent with Eq.~(\ref{eq:mtfix}) do indeed lead to $b-\tau$ mass
unification.
In this sense, our results are more general than the previously
cited analyses. We think it is likely that the approximate unification of
$\mb$ and $\mtau$ is telling us important physics, but we think it is premature
to draw conclusions from it.


\section{Electroweak Gauge Symmetry Breaking}\label{ewsb:sec}

One of the most remarkable features of the MSSM is a ``built-in'' mechanism
for dynamical electroweak symmetry breaking (EWSB)~\cite{gsymbreak}.
The renormalization group improved supersymmetric Higgs potential naturally
breaks $SU(2)\times U(1)_Y\rightarrow U(1)_{em}$ if the top quark Yukawa
coupling is sufficiently large compared to the gauge couplings.
As we outline below this
will allow us to reduce the number of free parameters in the theory and express
some GUT-scale free parameters in terms of more useful low-energy ones.

The tree level Higgs potential can be derived from
the expressions for $W$, Eq.~(\ref{spotential:eq}), the so-called D-terms,
and $\Lsoft$, Eq.~(\ref{lsoft:eq}):
\be
V_0 = m_1^2\vert H_d^0\vert^2 + m_2^2\vert H_u^0\vert^2 +
m_3^2(H_d^0H_u^0 +
h.c.) + \frac{g_1^2+g_2^2}{8}(\vert H_d^0\vert^2 - \vert
H_u^0\vert^2)^2,
\label{vtree:eq}
\ee
where $m_{1,2}^2\equiv m_{H_{d,u}}^2 + \mu^2, m_3^2\equiv B\mu$,
and the phases of the fields are chosen such that $m_3^2<0$.

Using the RGEs, one may define the renormalization group improved tree
level Higgs potential, $V_0(Q)$, at any scale $Q$. $V_0(Q)$ is understood
to be the tree level $V_0$ where the fields and coefficients have attained a
scale-dependence through their 1- or 2-loop RGEs.
However, as was emphasized in Ref.~\cite{grz}, in
general $V_0(Q)$ can
depend strongly on the energy scale at which it is evaluated. In
other words, minimizing $V_0(Q)$ at, say, $Q=\mz$ and again at some
slightly larger $Q$ may
lead to very different values of $v_d$, $v_u$, and therefore
$\tanb\equiv v_u/v_d$. This behavior is due to large radiative corrections
coming particularly from mass splitting in the $t-\stp$ system. If one knew
the scale $Q\sim O(\mz)$ at which these corrections were small, one could
safely minimize $V_0(Q)$ there. However, this scale in unknown {\it a priori}.
A much more satisfactory solution is achieved by minimizing the full one-loop
Higgs effective potential. The full Higgs potential can be
written as
\be
\vhiggs(Q)= V_0(Q) + \Delta V(Q),
\ee
where (see \eg, Ref.~\cite{sher})
\be
\delV(Q)=\frac{1}{64\pi^2}{\rm STr}\,M^4\left(\log\frac{M^2}{Q^2}-
\frac{3}{2}\right)
\ee
is the 1-loop contribution to $\vhiggs$ and
${\rm STr}\,f(M^2)\equiv\sum_j (-1)^{2j}(2j+1){\rm Tr}\,f(M^2)$
where $M$ and $j$ are the (field-dependent) mass and the
spin of a given state, and the
sum is over all states in the Lagrangian.

Electroweak symmetry-breaking can occur if the following
two conditions are met:
{\it (i)} $\vhiggs$ is bounded from below (\ie\ $m_1^2+m_2^2\geq 2\vert
m_3^2\vert$), and {\it (ii)} the minimum of $\vhiggs$ occurs at non-zero
field configurations (\ie\ $m_1^2m_2^2\leq m_3^4$). It was realized early
that, given a ``large'' top quark mass, EWSB could
be achieved radiatively~\cite{gsymbreak}. That is, despite taking $m_{H_d}^2,
m_{H_u}^2,\mu^2
>0$ at $\mgut$, requirement {\it(ii)} above can still be satisfied.
For a ``large'' $\mt\gsim 80\gev$, the running of $m_2^2$
is dominated by $h_t$, the top Yukawa. As the scale $Q$
decreases from the GUT scale, $m_2^2$ is driven
negative while $m_1^2$ and $\mu^2$ remain positive.

Minimization of $\vhiggs$ leads
to the system of equations:
\be
\frac{\partial\vhiggs}{\partial v_d^2} = m_1^2+m_3^2\tanb+
\onehalf \mz^2\cos 2\beta+ \Sigma_1 =0 \label{eq:dV11}
\ee
\be
\frac{\partial\vhiggs}{\partial v_u^2} = m_2^2+m_3^2\cot\beta-
\onehalf \mz^2\cos 2\beta+ \Sigma_2 =0 \label{eq:dV01}
\ee
where $\Sigma_{1,2}\equiv\partial\delV/\partial v_{d,u}^2$ and
all terms are implicitly $Q$-dependent.

Solving Eqs.~(\ref{eq:dV11}) and (\ref{eq:dV01}) one finds:
\be
\sin 2\beta(Q) = \frac{-2m_3^2(Q)}{\mu_1^2(Q)+\mu_2^2(Q)}  \label{eq:s2beta}
\ee
and
\be
\onehalf \mz^2(Q)=\frac{\mu_1^2(Q)-\mu_2^2(Q)\tan^2\beta(Q)}{\tan^2\beta(Q)-1}.
\label{eq:halfmz}
\ee
We have introduced two parameters:
\be
\mu^2_{1,2}(Q)\equiv m_{1,2}^2(Q) + \Sigma_{1,2}(Q)
= m_{H_{d,u}}^2(Q) + \mu^2(Q) + \Sigma_{1,2}(Q).
\ee

Examining Eqs.~(\ref{eq:dV11})--(\ref{eq:halfmz}), we find that, in fact, EWSB
can occur for any value of $\mt$ so long as $\mt>\mb$. In the limit
$\mt$ approaches $\mb$ (ignoring for now the contributions beyond the tree
level),
$\mu_2^2$ approaches $\mu_1^2$ from below, but is not driven negative as in the
large $\mt$ limit. Eq.~(\ref{eq:halfmz}) can now
only be satisfied as $\tanb$ approaches 1 from above. One concludes therefore
that radiative EWSB can occur for any $\mt>\mb$, though
small $\mt$ ($\lsim\mw$) would have required
$\tanb\simeq 1$. From a rough search of the parameter space we find that
although the condition $\mt\gsim\mw$ is always sufficient for EWSB
(assuming appropriate values for the other parameters), it is also
{\em necessary} in order to obtain values of $\tanb\gsim 2$.

It would be simplest if we could always minimize $\vhiggs(Q)$ at $Q=\mz$
because we know from experiment the value for $\mz(\mz)$ in
Eq.~(\ref{eq:halfmz}) above. In minimizing $V_0(Q)$, this would be
dangerous. But $\vhiggs(Q)$, unlike $V_0(Q)$,
is relatively stable with respect to $Q$, so that we can choose $Q=\mz$
with confidence.

The complete forms of $\Sigma_1$ and $\Sigma_2$ are
included in Ref.~\cite{anveff}. It has been
emphasized~\cite{anveff,casas} that the use of only the leading $t-\stp$
contributions to $\delV$ can be misleading due to
potentially large cancellations that can
occur with other terms that are not included.
Throughout our analysis, all contributions
to the complete 1-loop effective
potential have been included. Because use of the full potential
requires knowledge of the
complete SUSY spectrum, the iterative procedure that will be outlined in
Section~\ref{procedure:sec} is ideally suited for considering this issue.


\section{Supergravity-Based Constraints}\label{sugra:sec}

While the phenomenology of the MSSM is sometimes studied without referring to
its GUT-scale origin, we want to consider in this study a highly
constrained SUSY scenario
with as many well-motivated assumptions as possible. This will of course
enhance predictability for the ranges of parameters where
SUSY may be realized. (Later we can examine what modifications result from
relaxing assumptions.)

As we mentioned in Section~\ref{basics:sec}, a natural and often
considered approach is to couple the MSSM to minimal $N=1$ supergravity
from which the following set of assumptions emerges:
\begin{enumerate}
\item
{\em Common gaugino mass $\mhalf$:} The soft SUSY breaking gaugino mass terms
are equal to $\mhalf$ at $\mgut$,
\be
M_1(\mgut)=M_2(\mgut)=\mgluino(\mgut)\equiv\mhalf.
\label{mhalf:eq}
\ee
\item
{\em Common scalar mass $\mzero$:} The soft SUSY breaking scalar mass terms
contributing to the squark, slepton, and Higgs masses are equal
to $\mzero$ at $\mgut$,
\be
m_{\tildeQ}^2(\mgut)=m_{\tildeuc}^2(\mgut)=\cdots
=m_{H_d}^2(\mgut)=m_{H_u}^2(\mgut)\equiv\mzero^2.
\label{mzero:eq}
\ee
\item
{\em Common trilinear scalar coupling $\azero$:} The soft trilinear SUSY
breaking terms are all equal
to $\azero$ at $\mgut$,
\be
A_t(\mgut)=A_b(\mgut)=A_\tau(\mgut)=\cdots\equiv\azero.
\label{azero:eq}
\ee
\end{enumerate}

Through the RGEs of the MSSM, assumption~(\ref{mhalf:eq}) is often expressed
\bea
M_1 &=& {5\over3}\tan^2\theta_{\rm w}M_2\simeq0.5 M_2,\\ \label{monemtwo:eq}
M_2 &=& \frac{\alpha_2}{\alphas}\mgluino\simeq 0.3\mgluino,
\label{mtwomgluino:eq}
\eea
with $M_1$, $M_2$, and $\mgluino$ evaluated at the electroweak scale.
One also derives $\mhalf\simeq1.2M_2\simeq 0.36\mgluino$.

Assumptions~(\ref{mhalf:eq}) and~(\ref{mzero:eq}), in conjunction with SUSY and
the gauge structure, lead to the following expressions for the
masses of the sfermions (except for the third generation sfermions)
at the electroweak scale (see, \eg, Ref.~\cite{ibanez})
\be
m_{\widetilde\flr}^2 = m_f^2+\mzero^2 + b_{\widetilde\flr}\mhalf^2
\pm\mz^2\cos 2\beta\, [T_3^{\flr} - Q_{\flr}\sinsqthw],
\label{sfmass:eq}
\ee
where $\widetilde\flr$ is the left (right) sfermion corresponding to an
ordinary left (right) fermion, $T_3^{\flr}$ and $Q_{\flr}$ are the
third component of the weak isospin and the electric charge of the
corresponding fermion $f$,
and the coefficients $b$ can be expressed as
functions of the gauge couplings at $\mz$ and are $b\simeq6$ for squarks,
$\simeq0.5$ for left-sleptons, and $\simeq0.15$ for right-sleptons (see, \eg,
Ref.~\cite{sugradm}). Their exact values vary somewhat with different input
parameters.

While the assumptions~(\ref{mhalf:eq}),~(\ref{mzero:eq}), and~(\ref{azero:eq})
derive from theoretical speculations at the GUT scale, we want to stress that
some motivation for
assuming at least the common scalar mass is provided by experiment.
The near mass degeneracy in the $K^0-\bar K^0$ system implies
a near mass degeneracy
between $\widetilde s_L$ and $\widetilde d_L$~\cite{dgs}.
Similarly, slepton masses have to be strongly degenerate from
stringent bounds on $\mu\rightarrow e\gamma$~\cite{dgs}. It is thus sensible to
generalize this property to
all the mass terms, especially since there exists a well-motivated theoretical
framework providing it. Alternative
approaches exist~\cite{stringsoft,seiberg}, though we do not consider them in
this study.
We note that for almost all topics and applications only $A_t$ among the
trilinear
soft terms plays a role, so in practice we did not have to impose the
condition~(\ref{azero:eq}). The assumptions~(\ref{mhalf:eq}),~(\ref{mzero:eq})
will be easily tested
with any superpartner data.

Many past
analyses have also relied
on the further assumption that $\bzero=\azero-\mzero$ ($\bzero=B(\mgut)$,
\etc),
which follows from a restricted class
of SUGRA models. As has been shown in Ref.~\cite{giudiceroulet}, even if this
relation is present at tree level in the full theory,
it can be altered dramatically as heavy states are decoupled at $\mgut$.
We do not impose this constraint anywhere in the analysis.


\section{Procedure}\label{procedure:sec}

\subsection{Choice of Independent Parameters}\label{choice:sec}

After making the (SUGRA-inspired) reduction of the parameter space outlined
above, we are left with six ``fundamental'' input parameters at the GUT scale:
$\mzero$, $\mhalf$, $\azero$, $\bzero$, $\muzero$,
and $h_{t0}$. (In addition, we include all effects due to $\hbot$ and
$\htau$ in the analysis.)
However, not all of the parameters remain independent when we impose
radiative EWSB. Eq.~(\ref{eq:halfmz}) allows us to eliminate
$\mu^2(\mz)$ as a free parameter in favor
of $\mz^2$, though the sign of $\mu$ is still free.
Similarly, we can eliminate $B(\mz)$ in favor of $\tanb(\mz)$ via
Eq.~(\ref{eq:s2beta}).
Finally, given $\tanb$ and the RGEs we can replace $h_{t0}$ by $\mt$.
Table~\ref{params:table} summarizes our choices.

\begin{table}[b]
\centering
{\small
\begin{tabular}{|c|c|} \hline
	& $\mzero$, $\mhalf$, $\azero$, $\sgnmu$, $\mt$, $\tanb$ \\
Inputs  & $\mtau$, ($\mb$ or $\frac{\mb}{\mtau}|_{{}_{\mgut}}$) \\
        & $\alpha(\mz)$, $\sinsqthw(\mz)$  \\ \hline
        &  $\bzero$, $|\muzero|$ \\
        & ($\frac{\mb}{\mtau}|_{{}_{\mgut}}$ or $\mb$), $\alphas(\mz)$ \\
Outputs & $M_1$, $M_2$, masses and mixing angles of \\
        & gluinos, neutralinos, charginos, squarks, sleptons,\\
        & and Higgs bosons; $\abundchi$,
          BR$(b\rightarrow s\gamma)$, \etc  \\ \hline
\end{tabular}}
\caption{ Summary of input to and outputs from our analysis. Note that
the choice between $\mb$ and $\frac{\mb}{\mtau}|_{{}_{\mgut}}$ depends on
whether
we are testing the assumption of GUT-scale Yukawa unification as in
Section~\protect\ref{mbtau:sec} or requiring physically realistic bottom quark
masses
as in this Section and those that follow.}
\label{params:table}
\end{table}

This ``mixed'' set of input parameters,
$\mzero$, $\mhalf$,
$\azero$, ${\rm}\,\muzero$, $\tanb$, and
$\mt$,
has been commonly
used in the literature because of its technical convenience.
This convenience becomes apparent upon inspection of the system of
RGEs, in which $\mu$ and $B$ do not affect the running
of any of the other parameters
in the low-energy effective Lagrangian. Their
values at the weak scale may be
calculated from Eqs.~(\ref{eq:s2beta}) and
(\ref{eq:halfmz}), and run back up to $\mgut$
in order to determine $\muzero$ and $\bzero$.
The sign of $\mu$ is scale-independent.
Note that when we consider $\tanb$ in this analysis, we will always assume
$\tanb(Q)=\tanb(\mz)$ for all $Q\sim O(\mz)$; this is
well-motivated by the very slow running of $\tanb$ and the small range of
scales over which we consider the phenomenology of the MSSM,
and so introduces only negligible
errors.

There is another reason for the above
choice of input parameters. In some schemes, it is possible to
determine $\mt$ as an output. We feel,
however, that $\mt$ should be an input into any routine. Current LEP
data put strong constraints on
$\mt$, and direct discovery of the top quark at the
Tevatron may be forthcoming. Thus $\mt$ will soon
serve as a relatively well-known input parameter.
Therefore, analyses that
give $\mt$ as an output will not be efficient
in exploring the parameter space consistent
with a known $\mt$.

There is however a certain technical difficulty associated with using the
``mixed'' parametrization. Some
input parameters, like the Yukawa
couplings of the third generation, the gauge couplings,
and $\tanb$, are known or chosen at the $Z$-scale.
But others such as $\mzero$, $\mhalf,$ and $\azero$ are chosen at the GUT
scale. Furthermore, the two scales are mixed in the sense
that we must calculate the
values of $\mgut$ and $\agut$ through the running of the
low-energy values of the gauge
couplings. This running is in turn dependent on the
low-energy mass spectrum of the
SUSY particles, which depends most heavily on the values
of $\mzero$ and $\mhalf$ at the
GUT scale.
Therefore we employ an iterative numerical procedure that converges on a
consistent solution given all the input parameters.
We discuss it below.

\subsection{Running the RGEs}\label{running:sec}

We begin our numerical procedure at the electroweak scale, which we
take
to be $\mz$. This is an obvious choice since many experimental
quantities are now available at that scale.

At $Q=\mz$ we take as input the well-measured values of the $Z$
mass~\cite{lepreview},
\be
\mz=91.187\pm0.007\gev,
\label{mzmass:eq}
\ee
the electromagnetic coupling constant,
\be
\alpha(\mz)={1\over{127.9\pm0.1}},
\label{alphaem:eq}
\ee
and
the weak mixing angle,
$\sinsqthw(\mz)$, in the $\MS$ scheme. (The $\MS$ value of $\sinsqthw$
at the $Z$-pole is
defined so that
$\sinsqthw\cos^2\theta_{\rm w}\equiv(\pi\alpha/\sqrt{2}\,G_F)/
\mz^2(1-\Delta\widehat{r})$, where the radiative correction function
$\Delta\widehat{r}$ depends on both $\mt$ and $\mh$.)
The current world average for the weak mixing angle
is $\sinsqthw=0.2324\pm0.0008\pm0.0003$~\cite{lp1},
where the first error is due to
uncertainty in the value of $\mt$ and the second error is dominated by
the Higgs mass uncertainty. Because we take $\mt$ as
a known input parameter in this
analysis, the uncertainty due to top quark mass is
replaced by a functional dependence of $\sinsqthw$ on $\mt$~\cite{lp1}:
\be
\sinsqthw = 0.2324 -
1.03\times10^{-7}\left[\mt^2-
(138\gev)^2\right] \pm 0.0003 \label{eq:s2w}
\ee
One can see the dependence of the gauge couplings
on this parametrization in Table~\ref{sinsqthw:table},
where we have shown the values of $\alphas(\mz)$,
$\agut$, and $\mgut$ for several values of $\mt$,
using Eq.~(\ref{eq:s2w}), for a set of sample input parameters.
\begin{table}
\centering
{\small
\begin{tabular}{||c||c|c|c||} \hline
$\mt(\gev)$		& 120	& 145	& 170 	\\ \hline
$\sinsqthw$		& .2329	& .2322	& .2314	\\ \hline\hline
$\alphas(\mz)$		& .126	& .127	& .129 	\\ \hline
$\alphax$		& .0414	& .0413	& .0414	\\ \hline
$\mgut/10^{16}\gev$	& 1.76	& 1.94	& 2.26	\\ \hline
\end{tabular}}
\caption{ Typical effect of dependence of $\sinsqthw$ on
$\mt$ for $\mt=120$, $145$, $170\gev$ on $\alphas(\mz)$,
$\alphax$, and $\mgut$.
For this table we have chosen $\mzero=\mhalf=200\gev$,
$\tanb=5$, $\azero=0$, and $\mu>0$.}
\label{sinsqthw:table}
\end{table}
It is also interesting to note that this dependence of $\sinsqthw$ on
the top
quark mass leads to a strong dependence of $\alphas(\mz)$ on the top mass
as well.
For constant $\sinsqthw$, larger values of the top quark mass
tend to decrease the value of $\alphas(\mz)$ by about 3\% over the
allowed
range of $\mt$. However, due to the strong dependence of $\sinsqthw$
on $\mt$, the
value of $\alphas(\mz)$ actually increases by about 3\% over the same
range.

Besides the values of the gauge couplings at $Q=\mz$, one also needs
the
Yukawa couplings of the third generation of quarks and leptons
at $\mz$. The mass of the $\tau$ is now very precisely known,
$\mtau=1776.9\pm0.5\mev$~\cite{taumass}.
The mass of the $b$-quark, however, has a larger
uncertainty.
Following the analyses of Refs.~\cite{florida1,ty}
we take the central value of $\mbpole(\mbpole)$ to be
$4.9\gev$. In order to determine $h_b$ and $h_\tau$ at $Q=\mz$ we run
the gauge couplings $\alpha$ and $\alphas$ from their experimental
values at $Q=\mz$ down to the $b$- and $\tau$-mass scales
using 3-loop QCD and 2-loop QED RGEs~\cite{lowRGEs}.
At the mass thresholds, we translate~\cite{pole} the experimentally
measured pole masses to
the $\MS$ scheme and run these masses back up to the $Z$-scale.
Similarly we arrive at $h_t(\mz)$ by running gauge couplings up to the top
quark
mass threshold and then running $h_t$ back down to $\mz$.

Now we return to a careful treatment of the threshold corrections in
the running of the gauge couplings already mentioned in Section~\ref
{thresholds:sec}.
In the present analysis, all thresholds are handled as an intrinsic part of
the numerical routines. Because we determine the (running) mass of each
sparticle
at the scale $Q=m_i(Q)$ anyway, we can
simultaneously change the gauge coupling $\beta$-function coefficients
to reflect the coupling or decoupling of this particular state. We have
already argued in Section~\ref{thresholds:sec} that the RGEs must be run at
2-loops with correct 1-loop thresholds, which is what we do.
In Table~\ref{loops:table},
we demonstrate the importance of both these requirements. Notice in particular
that the net effect of the 2-loop running is to increase $\alphas(\mz)$ by
$O(10\%)$. Also notice that had we considered proton decay in this analysis,
we would have found that the proton lifetime coming from dimension-6 operators
increases when using 2-loop running instead of 1-loop
by a factor of $\sim 5$ since $\mgut$ has increased by $O(50\%)$ and the
lifetime scales as $\mgut^4$.
\begin{table}
\centering
{\small
\begin{tabular}{||c||c|c|c||c|c|c||} \hline
 ~ & \multicolumn{3}{c||}{Example 1} & \multicolumn{3}{c||}{Example 2} \\
\hline
   & 1-loop & 2-loop & $\msusyeff$ & 1-loop & 2-loop & $\msusyeff$ \\
\hline\hline
$\alphas(\mz)$		& .117	& .129	& .117 + 2-loops & .111 & .121
& .111 + 2-loops  \\ \hline
$\alphax$		& .0404	& .0422	& .0456 & .0380 & .0394 & .0417 \\
\hline
$\mgut/10^{16}\gev$	& 1.50	& 2.37	& 4.30  & 0.79 & 1.18 & 1.81 \\
\hline
\end{tabular}}
\caption{ Typical effect of dependence of $\alphas(\mz)$, $\alphax$,
and $\mgut$ on 1-loop running, 2-loop running, and 2-loop running with the
``effective scale'' of Eq.~(\protect\ref{msusyeff:eq}) for two spectra of SUSY
particles. For Case~1, we take
$\mzero=\mhalf=100\gev$, $\tanb=5$, $\mt=145\gev$, $\azero=0$, and $\mu>0$.
Case~2 is the same as Case~1 except $\mzero=\mhalf=1\tev$.
For the two cases,
$\msusyeff=13, 177\gev$ respectively. Recall that $\msusyeff$ is defined
to reproduce the 1-loop value for $\alphas(\mz)$. We calculate the values of
$\alphax$ and
$\mgut$ for $\msusyeff$ as in Ref.~\protect\cite{cpw}.}
\label{loops:table}
\end{table}

When running the gauge coupling RGEs, we follow
the decoupling prescription outlined in Eqs.~(\ref{eq:b1})--(\ref{eq:b3}).
However, there are some minor simplifications and ambiguities to
consider~\cite{ekngutcorrs}. First, we decouple all higgsinos at the common
scale $Q=\mu(Q)$, binos
at $Q=M_1(Q)$, winos at $Q=M_2(Q)$, and the second Higgs doublet at
$Q=m_A(Q)$. For the top quark, one could either choose to decouple it at
its mass threshold, or simply at $\mz$; numerically either procedure is
essentially equivalent. One other ambiguity in the 1-loop RGEs arises for
weak isodoublets decoupling from $\beta_2$. Here, because
they will always appear in $T_3=\pm\onehalf$
pairs in the loops, we only couple the doublet when the scale
is larger than the heavier member of the doublet.
This can be seen in Eq.~(\ref{eq:b2}). At 2-loops
many such ambiguities arise; however, the effects of individual thresholds
in the 2-loop RGEs are of higher order and can be safely
ignored. Therefore we have changed the 2-loop coefficients with a single
threshold at $Q=\mhalf$ above which we use MSSM 2-loop coefficients and
below which we use those of a two-Higgs doublet SM.
We have checked that dramatically varying the scale $Q$ at
which the 2-loop coefficients are changed from their SM values to the
SUSY ones causes typically an $O(2\%)$ variation in $\alphas(\mz)$
($\Delta\alphas(\mz)\lsim0.002$).
Thus, we feel that our approximation is justified.

It is important to reiterate that we only decouple states in the running of the
gauge couplings. This decoupling is necessary in order to determine realistic
values for $\alphas(\mz)$.
However, were one to decouple states, say, from the soft mass RGEs,
then one would need to reconsider the effective Lagrangian and matching
conditions at scales below each threshold, where this Lagrangian,
its couplings, and their RGEs would no longer be supersymmetric.
By minimizing the 1-loop effective potential with all states included down to
$Q=\mz$, we effectively include the contributions from their decoupling.
Therefore, in all RGEs other than those of the gauge couplings, we have
left all states coupled down to $Q=\mz$ where we minimize the full 1-loop
effective potential.

Once the boundary conditions at the GUT scale have been set,
we run the RGEs of the  system in order
to determine the value of
a parameter at any scale $Q$ below $\mgut$.
The RGEs for minimal SUSY have
appeared in numerous places in the literature, including
Refs.~\cite{susyRGEs,ez};
we follow essentially the conventions of Ref.~\cite{ez}.
Though various authors have offered semi-analytic,
approximate solutions to the full
set of RGEs under various simplifying assumptions, a full
analysis of the parameter space
requires that the RGEs be solved numerically, given the level of accuracy
that we are maintaining.

The procedure that we adopt essentially consists of repeatedly running
the RGEs between $Q=\mz$ and $Q=\mgut$ until a self-consistent solution has
been isolated.

In the first iteration for any given set of input
parameters, an approximate SUSY
spectrum is generated. The six RGEs of the gauge and Yukawa couplings,
are simultaneously run up first to the
GUT scale using the method of Runge-Kutta.
We run the gauge couplings in the SUSY-consistent $\DR$ scheme as opposed
to the $\MS$ scheme, and so we impose the
matching condition for the two schemes at $Q=\mz$~\cite{ekngutcorrs}.
(The net effect of the scheme change is less than 1\%
however~\cite{lp1,ekngutcorrs}.)
Running up, we define $\mgut$ as that
point at which $\ai(\mgut)=\aii(\mgut)\equiv\agut$.
We then set $\alphas(\mgut)=\agut$.
All scalar masses are set equal to $\mzero$, all gaugino masses to $\mhalf$,
and all $A$-parameters to $\azero$.

The RGEs for all the twenty-six running parameters (the gauge and
Yukawa couplings, the $\mu$-parameter, and the soft mass terms)
are run back from $Q=\mgut$ down to $Q=\mz$.
For the gauge and Yukawa couplings, 2-loop RGEs with thresholds are
used throughout, while 1-loop
RGEs are used for the SUSY soft mass parameters.
Along the way, we decouple any particle $i$ in the spectrum from the
gauge coupling RGEs at the scale $Q=m_i(Q)$. As described earlier, thresholds
in the 1-loop gauge coupling RGEs are used to account for the effects of
the decoupling of the various sparticles at masses greater than $\mz$. At
$Q=\mz$ a value for $\alphas(\mz)$ is found
consistent with unification assumptions, and the full 1-loop
effective scalar potential is
minimized in order to determine the values of $\mu^2(\mz)$ and
$B(\mz)$ that produce proper EWSB.
On the next iteration when the entire set of parameters is again run
up from $Q=\mz$ to a newly determined $\mgut$, the parameters $\mu$ and $B$
will also run, providing their
corresponding values at the GUT scale.

This entire procedure is repeated several times, terminating
only after changes in the
solutions to the RGEs are small compared to the values
themselves or to the experimental
errors, whichever are relevant. Each iteration provides
a more precise spectrum of
sparticles, which in turn provides more precise running of the gauge and
Yukawa couplings. We find that the whole procedure is
extremely stable, usually converging to a
solution in just a few iterations.

In Figure~\ref{running:fig}, we give an example of the running
of various sparticle masses from the GUT scale down to the electroweak scale.
Notice that the mass of the Higgs that couples to the top quark is driven
imaginary (\ie, its mass-squared is driven negative)
at scales of $O(1\tev)$, signally the onset of EWSB. This is shown in the plot
as the mass itself going ``negative'' for convenience of presentation.

When the program has isolated a solution, we have as our output all sparticle
masses and mixings valid to 1-loop, Higgs masses which include all third
generation contributions to the
1-loop radiative corrections~\cite{radcorrs},
$\alphas(\mz)$, $\agut$, and $\mgut$ valid to 2-loops, and
the GUT-scale parameters $\bzero$ and $\muzero$.


\section{Constraints}\label{constraints:sec}

In applying the numerical procedure described in the previous section
we have required the gauge coupling to unify,
and from the input values of $\alpha$, Eq.~(\ref{alphaem:eq}), and $\sinsqthw$,
Eq.~(\ref{eq:s2w}), obtained a range of $\alphas(\mz)$ as a function of
independent parameters. We have also demanded proper
EWSB yielding the experimentally measured value of $\mz$. We have
parametrized the many mass parameters of the MSSM in the usual way,
assuming common gaugino and scalar masses and the $A$ parameters,
Eqs.~(\ref{mhalf:eq}),~(\ref{mzero:eq}), and~(\ref{azero:eq}), as implied by
minimal SUGRA.
Before we present our results in the next section, we now list and briefly
elaborate on several other constraints
that we will impose on the output of our numerical analysis. As we explained in
Section~\ref{mbtau:sec}, we do not impose the condition
$\mb=\mtau$ at the GUT scale because the resulting bottom quark mass
is likely to be very sensitive to the threshold corrections at $\mgut$,
which we cannot include without selecting a specific GUT model. Without such
corrections we obtain the values of $\mbpole$ about 20\% above the
current experimental range, except for very large $\mt$.

\subsection{Limits from Experimental Searches}\label{limits:sec}

LEP experiments have placed lower limits on the chargino mass of about
47\gev, and on the charged slepton, sneutrino,
and squark masses of about 43\gev~\cite{riles}. The lightest stop
mass bound is dependent on the left-right stop mixing, which can reduce
its coupling to the $Z$-boson. DELPHI~\cite{delphistop} has excluded
$\mstopone$ below about 45\gev, except for a rather tiny range of the mixing
angle which allows $\mstopone>37\gev$.

Placing an experimental lower limit on the masses of the Higgs bosons is in
principle more
complicated since either $\hl$ or $\ha$, or both, can be light, and because of
potentially sizable radiative corrections to their masses due to
the heavy top quark. Assuming reasonable ranges of value for $\mt$,
$\mstopq$, and $\tanb>1$ the bounds $\mhl>44\gev$ and $\mha>21\gev$ have
been derived by ALEPH~\cite{riles}. Other LEP experiments obtained similar
limits. However, once we impose the unification and EWSB conditions, we find
that $\hl$ couplings are very SM-like ($\sin^2(\beta-\alpha)\approx1$) so that
in practice the LEP limit of
about 62\gev~\cite{higgs7} for the SM Higgs applies to $\hl$ as well.
For related reasons, $\ha$ is always heavier than $\mz$ for us, so that the
LEP limits on $\ha$ place no serious bound.

Lower mass bounds on the squarks and gluino have been reported at
126\gev\ and 141\gev~\cite{abe}, respectively, assuming no cascade decays.
By including cascade decays one can reduce those bounds by some 20\gev\
or more~\cite{btcascade}. The squark masses could become even as light as
allowed by LEP
if $\mgluino$ becomes large. All squark and gluino bounds are very model
dependent.

LEP experiments alone cannot place
a lower bound on the mass of the lightest
neutralino because its coupling to the $Z$ can be strongly suppressed and it is
not directly detectable.
It is only
by combining LEP direct chargino and neutralino searches
with indirect ($Z$-line shape)
searches and with the lower bound on $\mgluino$ from the Tevatron
(via Eq.~(\ref{mtwomgluino:eq}))
that a bound $\mchi\gsim18\gev$~\cite{chilowerlimit} can be derived
for any $\tanb>1$.

\subsection{$b\to s\gamma$}\label{brbsg:sec}

Recently, CLEO has reported an upper bound on
BR($b\to s\gamma )<5.4\times 10^{-4}$~\cite{cleo7}.
A central value ($3.5\times 10^{-4}$) and a lower limit ($1.5\times 10^{-4}$)
is obtained from the detection of $B\to K^*\gamma$~\cite{cleo7} and
assuming that the ratio of BR($B\to K^*\gamma$) to BR($b\to s\gamma$)
is 15\%~\cite{cornell7}. In addition to the SM contribution, SUSY allows for
one-loop diagrams with the exchange of the charged Higgs and the charginos and
neutralinos~\cite{bertolini7,barbieri7,hewett7,oshima7,garisto7}.
We calculate BR($b\to s\gamma$) with the formulas of Ref.~\cite{barbieri7}.
These use QCD corrections that are less accurate than has been done
for the SM case recently~\cite{adel7}.
We are in the process of combining our results
with those of Ref.~\cite{adel7} to obtain improved
QCD corrected CMSSM predictions.

In Section~\ref{results:sec} we will apply the {\em upper} bound
BR($b\to s\gamma )<5.4\times 10^{-4}$.
In Section~\ref{bsg:sec} we will present the {\em predictions} of
CMSSM for BR($b\to s\gamma$) and show that the range favored by CMSSM naturally
falls into the range resulting from the CLEO analysis. We will also show that
the claims~\cite{hewett7} of a stringent bound on the charged Higgs
mass are too strong in the CMSSM.

\subsection{Color and Charge Breaking}\label{colorsub:sec}

In the MSSM the Higgs potential automatically conserves color and charge but
the same is not necessarily true with the full scalar
potential.
If one wishes to determine the form of the global minima, one must
numerically search for all local minima of the full scalar potential,
including  the charged and colored states,
and determine the broken symmetries associated with each.
This is outside the realm of the study we are reporting on here and so we
only demand that the $({\rm mass})^2$ of any charged or colored mass eigenstate
remains positive. In fact, this will be an
important constraint in some regions of the parameter space
(especially for large $\azero$) where the lighter stop $({\rm mass})^2$
can become negative due to a large $\stopl-\stopr$ mass splitting.

It is sometimes stated in the literature that a necessary condition for
avoiding color-breaking~\cite{savoy} is to demand that
$|\azero|/\mzero<3$. However, as pointed out by Ref.~\cite{ghs}, this condition
is really neither sufficient nor necessary. Therefore, we consider values
for $|\azero|/\mzero$ slightly larger than three. Knowing that
minimization of the
full scalar potential may lead to color- or charge-breaking minima for such
large
ratios, the constraints coming from our analysis can only be strengthened by
a full treatment of this color/charge-breaking.

Nonetheless, as will be discussed in the next Section, $m^2_{\stopone}$
never goes negative for smaller values of $\azero$ ($|\azero|/\mzero\lsim1$)
simply because the $\stopl-\stopr$ mass splitting is dominated by
$\azero$.

\subsection{Lightest Neutralino as the LSP and DM Candidate}
\label{lsp:sec}

In the absence of $R$-parity breaking the LSP remains absolutely stable.
Depending on its nature it may have to face potentially tight
cosmological constraints which we will discuss below. It will also have
important
experimental consequences for possible SUSY signatures. In the MSSM, any of the
superpartners could in principle be the LSP because their masses are virtually
unrelated. In the CMSSM the picture is very different: the masses of the
superpartners are highly correlated.
These relations are determined by the assumptions of
Eqs.~(\ref{mhalf:eq})--(\ref{azero:eq})
and lead to a hierarchy among the sparticle masses.
As a result there are very few possible candidates for the LSP.
Typically it is the lightest of the four neutralinos that comes out to be the
LSP, and it has been usually favored in most phenomenological and cosmological
studies. However, for some combinations of parameters some other sparticle,
like the stop, the stau, or the sneutrino can
be the LSP. Each of the resulting types of the LSP must meet
cosmological constraints.

As we have already discussed in Section~\ref{colorsub:sec}, due to a large mass
splitting in the $\stopl-\stopr$
sector, the lighter stop mass eigenstate ($\stopone$) may in certain
cases  become very light. In fact,
one may even encounter $m^2_{\stopone}<0$. On the other hand, the lighter stau
sometimes becomes the LSP.
As concerns the sneutrino, after we apply experimental limits and
reject unphysical cases, we never find it to be the
LSP.

\subsubsection{Neutral LSP}\label{neutrallsp:sec}

It would be difficult to imagine that an electrically charged or colored
massive stable particle, like the stau or the stop, could exist in any
meaningful amount in the Universe~\cite{kt,glashow}. If it did, it would
interact with
photons and become detectable. It would also interact with ordinary matter and
dissipate its energy thus falling towards the cores of galaxies. It would form
stable isotopes of chemical elements. For these, and other, reasons, only
electrically neutral and colorless particles are believed to be able to exist
in the Universe in the form of dark matter~\cite{kt,glashow}. We will
therefore reject those regions of the parameter space where either the stop or
stau are the LSP.
In the rest of the study we will only deal with the neutralino as the LSP.

\subsubsection{Neutralino Relic Abundance}\label{neut:sec}

Any stable (or meta-stable) species predicted by theory would contribute to the
total mass-energy of the Universe. A relic abundance is usually expressed as
the ratio of the particle's relic density to the critical density
$\rhocrit\equiv{3H_0^2}/{8\pi G}=1.9\times10^{-29} (h_0^2) g/cm^3$
\be
\Omega_\chi\equiv{\rho_\chi\over\rhocrit}
\label{omega:eq}
\ee
where $\rhocrit$ corresponds to the flat Universe
and $h_0$ is the present value of the
Hubble parameter $H_0$ in units 100~km/s/Mpc ($h_0={H_0\over{100\,km/s/Mpc}}$).
Current estimates only require
$0.4\lsim h_0\lsim1$~\cite{kt}.

A supersymmetric LSP, being stable,
cannot decay on its own but can pair-annihilate into ordinary matter.
Its relic abundance $\abundchi$ is inversely proportional to the LSP
annihilation cross section and thus depends on the masses and couplings of the
final and exchanged particles.
In calculating the neutralino relic abundance we include {\em all} the relevant
LSP pair-annihilation channels into ordinary matter that are kinematically
allowed. Lighter $\chi$s
annihilate only (except for rare radiative processes) into
pairs of ordinary fermions via the exchange of the $Z$ and the Higgs
bosons, and the respective sfermions. (We do not include final state gluons
since the relevant cross section has been shown to be relatively insignificant
in calculating the relic abundance in the early Universe~\cite{djkn}.) As
$\mchi$ grows new final
states open up: pairs of Higgs bosons, gauge and Higgs bosons, $ZZ$ and
$WW$, and $t\bar t$, all of which we include in our analysis.
The actual procedure of calculating the relic abundance is quite involved and
has been
adequately described elsewhere (see, \eg, Refs.~\cite{kt,swo,mydmreview}).
We use the technique developed
in Ref.~\cite{swo} which allows for a reliable (except near poles and
thresholds) computation of the thermally
averaged annihilation cross section in the non-relativistic limit and
integration of the Boltzmann equation. This technique is applicable to
calculating the relic abundance in most of the parameter space.

As was first pointed out in Ref.~\cite{kani}, and rediscovered and elaborated
by Griest and Seckel~\cite{gs}, special care must be applied to calculating the
relic abundance near the poles of exchanged particles and when new mass
thresholds become kinematically accessible.
In particular, proper treatment of narrow poles has been provided in
Refs.~\cite{lnz:dm,andm,gondolo} and it was
shown that standard techniques may lead to errors reaching even
two or three orders of magnitude in the vicinity of a pole. This is especially
true for the lightest Higgs because the width of $\hl$ is extremely narrow, and
also near the $Z$-boson pole where the effective coupling
is somewhat stronger. We find that the regions of the parameter space where our
(standard) calculation fails
are relatively small albeit non-negligible. In presenting our results
in the next Section we will therefore point out those regions where the
presented results for the neutralino relic abundance are not
trustworthy. It has been argued in Refs.~\cite{lnz:dm,andm} that the regions
where the $\hl$ and $Z$-poles dominate are favored by current limits on
the proton decay in the SUSY $SU(5)$ model. Since we do not select $SU(5)$ as a
GUT symmetry, nor view it as particularly attractive, at this
point we choose not to pay special attention to calculating
the relic abundance near the poles. We will comment on these effects
in discussing results.

\subsubsection{Age of the Universe}\label{age:sec}

In the Standard Cosmological Model the age of the Universe depends on
the total relic abundance $\Omega_{\rm TOT}$. Conversely, estimates of the
Universe's age
place a constraint on $\Omega_\chi<\Omega_{\rm TOT}$. A conservative
assumption that
the Universe is at least 10 billion years old (and $h_0>0.4$)
leads to~\cite{kt}
\be
\Omega_\chi h^2_0\lsim 1.
\label{tenbillionyrs:eq}
\ee
If the age of the Universe is at least 15 billion years,
as many currently believe, then the bound~(\ref{tenbillionyrs:eq}) becomes much
stronger:  $\Omega_\chi h^2_0\lsim 0.25$~\cite{kt}. This is because an older
Universe corresponds to a smaller expansion rate
$h_0$. No stable particle can
contribute to $\Omega_{\rm TOT}$ more than is allowed by at least the
bound~(\ref{tenbillionyrs:eq}) without distorting the Universe's evolution.
This constraint is independent of the nature (or even existence) of dark matter
(DM) in the Universe.
The bound of Eq.~(\ref{tenbillionyrs:eq}) must be satisfied for any choice of
free
parameters and, as we
will see in the next Section, it provides a very strong constraint on the
parameter space.

\subsubsection{Dark Matter}\label{dm:sec}

The visible matter in the Universe accounts for about $1\%$ of the
critical density. There is at present abundant evidence for the existence of
significant amounts of dark matter in galactic halos
($\Omega\sim0.1$) and in clusters of galaxies ($\Omega\gsim0.2$)~\cite{kt}.
Big Bang nucleosynthesis (BBN) constrains the allowed range of baryonic
matter in the Universe to the range $0.02<\Omega_B<0.11$~\cite{bbn}
(and more recently $\Omega_B\approx 0.05$; see the second paper of
Ref.~\cite{bbn}).
The value
$\Omega_{\rm TOT}=1$ is strongly preferred by theory since it
is predicted by the models of
cosmic inflation and is the only stable value for
Friedmann-Robertson-Walker
models. Values of $\Omega_{\rm TOT}$ larger than those ``directly'' observed
are
also strongly supported by most models of large structure formation.
This, along with estimates given above, implies that: {\em (i)} most
baryonic matter in the Universe is invisible to us, and
{\em (ii)} already
in halos of galaxies one might need
a substantial amount of non-baryonic DM.
If $\Omega_{\rm TOT}=1$ then most (about 95\%) of the matter in
the Universe is non-baryonic and dark.
Current estimates of $h_0$ give, for $\Omega_{\rm TOT}=1$,
$0.5\lsim h_0\lsim0.7$ (the upper bound coming from
assuming the age of the Universe above 10 billion years),
in which case one expects
$0.25\lsim\Omega_{\rm TOT} h^2_0\lsim0.5$.
While it is not unlikely that the galactic halos consist to a large degree of
various extended MACHO-type objects\footnote{Recently, a few candidate events
for MACHO's with mass $\sim 0.1M_{\odot}$ have been reported by microlensing
experiments~\cite{machoevents} thus implying that some sort of small stars
comprise a significant component of the halo of our Galaxy. We note that, with
the present efficiency, this discovery does not, and will not for the next
several years, be
able to eliminate other kinds of candidates for the dark matter~\cite{kim}.
}
(like Jupiters, brown dwarfs, \etc),
it would be very hard to believe that such objects could fill out the whole
Universe without condensing into galaxies. This, along with
the bound on baryonic matter provided by BBN has led to a widely
accepted hypothesis that the bulk of DM in the Universe
consists of some kind of weakly interacting massive particles (WIMPs).
The relic abundance of the lightest neutralino $\chi$ (most naturally of
bino-type~\cite{chiasdm}) often comes out
to be in the desired range thus making it one of the best candidates for
DM~\cite{ehnos}. Being nonrelativistic, it falls into the category of
cold DM (CDM) which has been favored by models of large structure formation,
in contrast to hot DM (HDM), like light neutrinos. In a purely CDM scenario
one assumes that the LSP dominates the mass of the Universe, leading roughly to
\be
\label{abundrangecdm}
0.25\lsim\abundchi\lsim0.5.\ \ \ \ \ \ \ \ \ ({\rm CDM})
\ee

Motivated by the theoretical expectation that SUSY GUT theories will also have
massive neutrinos, and phenomenologically by the result that (in the aftermath
of COBE)
a mixed CDM+HDM picture (MDM) seems to fit the astrophysical data
better~\cite{mdm} than the pure CDM model, we also consider a smaller value of
$\Omega_\chi$. In the mixed scenario one assumes about
30\% of HDM (like light neutrinos with $m_\nu\simeq 6\ev$) and about
65\% of CDM (bino-like neutralino), with baryons contributing
the remaining 5\%.
In this case the favored range for $\abundchi$ is approximately given by
\be
\label{abundrangemdm}
0.16\lsim\abundchi\lsim0.33.\ \ \ \ \ \ \ \ \ ({\rm MDM})
\ee

(Strictly speaking, in the MSSM the neutrinos are massless and as such
could not constitute interesting HDM. But it is straightforward to extend the
model to include right handed neutrinos (and their sneutrino partners)
and give them mass terms. We don't expect this extension to sizably
modify the running of all the other parameters of the MSSM. It is with this
implicit assumption that we will
apply the range given by~(\ref{abundrangemdm}) in analyzing the resulting
implications for SUSY searches.)

Both scenarios assume a significant amount of LSP DM.
The sneutrino, an early candidate~\cite{snucandidate} for DM, is now
strongly disfavored.
After the LEP experiments have placed
a limit on its mass $m_{\tildenu}> 43\gev$, its relic abundance can now only be
negligibly small ($\Omega_{\tildenu}\sim 10^{-3}$).
We thus find it remarkable that we never find the sneutrino to be the LSP. Had
it been
the LSP instead of the neutralino in most of the parameter
space then the CMSSM would not have provided a viable candidate for
the DM problem!


\section{Results}\label{results:sec}

We now proceed to discuss the numerical results obtained by using the
procedure for generating low-energy output described in
Section~\ref{procedure:sec}. We will first analyze the impact of several
experimental, theoretical, and cosmological constraints on
the parameter space. Next, we will focus on the region of the model's
parameter
space consistent with all the adopted constraints and discuss
the resulting consequences for the value of $\alphas(\mz)$, the mass
spectra of the Higgs and supersymmetric particles, and other predictions.

\subsection{Effect of Constraints}\label{effectsresults:sec}

We have generated a large set of solutions for a broad range of input
parameters. We explore wide ranges of both $\mhalf$ and $\mzero$, each
between $50\gev$ and approximately $3\tev$ in 22 logarithmic steps, for
discrete values of $\mt=120, 145, 170\gev$, $\tanb=1.1$, 1.5, 3, 5, 10, 15, 20,
30, 40, 50, and $\azero/\mzero$ between -3.5 and 3.5 in increments of 0.5. We
also consider both signs of $\muzero$.

The choice of a logarithmic scale for $\mhalf$ and $\mzero$ is technically
motivated. We are interested most particularly in lower
values of the soft masses where the fine-tuning reintroduced by SUSY breaking
is smallest and where we can expect currently planned facilities to best
probe the parameter space. Likewise, the difference between $\tanb=1.5$ and
$\tanb=3$ is more significant than that between $\tanb=40$ and $\tanb=50$.
Lastly, the scaling of $\azero$ with $\mzero$ is motivated by SUGRA, with
bounds motivated by the fear of color-breaking global minima for large
$\azero$.

Mass
scales above $1\tev$ may seem unnatural but we also wish to explore the
asymptotic behavior of our results.
For the top mass, the three representative
values, $\mt=120$, $145$, $170\gev$, help us sample
the whole region of top mass preferred by the analysis of the LEP data.
We pay particular attention to the middle value, $\mt=145\gev$, as being
favored by LEP (when one includes a light Higgs as required by SUSY)
and perhaps by the cross section for candidates from the Tevatron.
For $\tanb$ we sample a spectrum of values over the entire region consistent
with the Yukawa couplings of the third generation remaining perturbative
all the way up to the
scale of unification. Values of
$\tanb\lsim 1.1$ become difficult to study due to a dangerous cancellation in
Eq.~(\ref{eq:halfmz}) and because we are close to the perturbative limit of the
top Yukawa; values of $\tanb\gsim 50$ become difficult because the bottom and
$\tau$ Yukawa couplings are likewise close to their perturbative limits.

Overall, we explore well over
100,000 combinations, each representing a unique point in the space of
$\mt$, $\tanb$,
$\mhalf$, $\mzero$, $\azero$, and $\sgnmu$.
We present some representative solutions  in
Figs.~\ref{envsone:fig}-\ref{casethree:fig} in the plane ($\mhalf,\mzero$). In
this Section we focus mostly on the case $\mt=145\gev$ and
several representative choices of $\tanb$ and $\azero$. We will also display
the dependence on $\mt$ below.

\subsubsection{Constraints from Experimental Searches}\label{exptresults:sec}

As we can see from Figs.~\ref{envsone:fig}-\ref{casethree:fig},
at present the regions of the
plane ($\mhalf,\mzero$) excluded by direct and indirect searches for
SUSY at LEP and FNAL (see Section~\ref{limits:sec}) are
limited at best. The strongest direct constraints on $\mhalf$ come from
$\mcharone>47\gev$ and/or the CDF gluino mass bound. Assuming
$\mgluino>141\gev$, and neglecting cascade decays, corresponds roughly
to $\mhalf\gsim50\gev$.
On the other hand, in general
there is no lower bound on $\mzero$ except for small $\mhalf$ from the
experimental lower bounds on the slepton and squark masses. As an example, we
present in Fig.~\ref{sneutrino:fig} the region of the plane ($\mhalf,\mzero$)
ruled out by the LEP bound $m_\sneutrino>43\gev$ for two extreme values of
$\tanb$ and by $\mgluino>141\gev$.
For $\mhalf\gg\mzero$ the exact
value of $\mzero$ becomes unimportant, as $\mhalf$ will come to dominate
the values of all
masses and will dictate how EWSB occurs. Though for the major
portions of this study we have taken $\mzero\geq 50\gev$, we have explored
the regions of much
lower $\mzero$ and found
nothing to change our conclusions as reported in
Sections~\ref{results:sec} and~\ref{implications:sec}.

In addition, we find some regions where the lighter stop mass
becomes smaller than the current experimental bound of about $37\gev$
and quickly becomes tachyonic as will be discussed below.

\subsubsection{Constraints from $b\ra s\gamma$}\label{bsgresults:sec}

Following Section~\ref{brbsg:sec}, we apply the upper bound
BR($b\to s\gamma)<5.4\times 10^{-4}$. Interestingly,
this bound is often quite important and particularly probes regions
of small to moderate $\mhalf$ and $\mzero$
(Figs.~\ref{envsone:fig}--\ref{envsthree:fig}). As
$\mhalf$ and $\mzero$ grow, BR($b\to s\gamma)$ tends to decrease
and produce the range of values consistent with CLEO for a wide range
of parameters as will be shown in Section~\ref{bsg:sec}.

\subsubsection{Constraints from and on $\alphas(\mz)$}\label{alphasresults:sec}

As we can see from Figs.~\ref{caseone:fig}-\ref{casethree:fig}, the values of
$\alphas(\mz)$ resulting
from our analysis generally fall into the experimentally allowed range. LEP
event shape measurements alone give
$\alphas(\mz)=0.123\pm0.006$~\cite{lepreview}
while other LEP analyses and low-energy
experiments typically yield somewhat lower ranges leading to the world-average
$\alphas(\mz)=0.120\pm0.006\pm0.002$~\cite{lepreview}.
(We note, however, that much smaller values of $\alphas(\mz)=0.107\pm0.003$
have been derived in Ref.~\cite{kramer}.)
We find that $\alphas(\mz)$ generally decreases with growing $\mhalf$ and
$\mzero$, and increases with $\mt$ (see Table~\ref{sinsqthw:table} and
Section~\ref{running:sec}).
Since small $\mhalf$ and $\mzero$ are excluded
by some experimental constraints (Section~\ref{exptresults:sec}),
we find $\alphas(\mz)\lsim 0.133$, including the range of very small $\mzero$.
This is a significant constraint on the entire picture and an important
prediction.
No interesting upper bound on the plane
($\mhalf,\mzero$) can be derived from a lower bound on $\alphas(\mz)$ because
$\alphas(\mz)$ decreases very slowly and reaches 0.110 for $\mhalf$ and/or
$\mzero$ in the range of tens of $\tev$.
Keeping SUSY masses below about $1\tev$ provides a lower bound
$\alphas(\mz)\gsim0.119$, while requiring no fine-tuning ($f\leq50$, see
Section~\ref{ftresult:sub}) gives $\alphas(\mz)\gsim0.118$.
Clearly, as the graphs show, larger values of $\alphas(\mz)$ are favored by
low-energy SUSY. Finally, if $\alphas(\mz)$ comes out much smaller (as
claimed in Ref.~\cite{kramer}), one may have to {\em necessarily} include
GUT-scale corrections to the running of the gauge couplings
in order to possibly remain in the theoretically favored region of
low-energy SUSY below a few~\tev. This may actually come out to be
a useful tool in an attempt to discriminate between different GUT scenarios.

\subsubsection{Constraints from EWSB}\label{ewsbresults:sec}

Proper EWSB is not automatic and requiring it places additional strong
constraints on the allowed combinations of parameters.
As can be seen in Figs.~\ref{envsone:fig}-\ref{envsthree:fig}, this constraint
excludes significant
regions in the upper left-corner ($\mzero\gg\mhalf$) of the plane
($\mhalf,\mzero$), unless
$\tanb$ is close to one or $\azero$ is larger and negative.
For
$\muzero>0$ there are additional regions in the lower right-hand
corner ($\mhalf\gg\mzero$) of the plane ($\mhalf,\mzero$) which are also
excluded for larger values of $\tanb$. This is because the full 1-loop
effective potential has become unbounded from below in those regions.

\subsubsection{Constraints from Avoiding Color
Breaking}\label{colorresults:sec}

As we said above, sometimes $\mstopone^2$ becomes negative. As one can see from
the presented
figures, this usually happens roughly for $\mzero\gsim\mhalf$ for rather
large values of $|\azero|$ (see symbol ``L" in these areas in, \eg,
Figs.~\ref{envsone:fig}c-d). The regions where $\mstopone^2<0$ always grow with
increasing $\tanb$. More specifically, for $\muzero<0$, color breaking occurs
when $\azero/\mzero\lsim-2$ for the whole range of $\tanb$, and also to some
extent
for $\azero/\mzero\gsim3$ and large $\tanb$. For the smaller values of $\azero$
$\mstopone^2$ is always positive, as expected.
For $\muzero>0$ the situation is generally similar for a reversed sign of
$\azero$.

\subsubsection{Constraints from Neutralino LSP}\label{lspresults:sec}

As we have argued in Section~\ref{neutrallsp:sec}, only the lightest
neutralino LSP remains a viable candidate for DM. On the other hand,
for $\mhalf\gg\mzero$ we invariably find that  the lighter stau
is the LSP, and not the neutralino, as one can see in
Figs.~\ref{envsone:fig}--\ref{envsthree:fig}. This is
expected since the mass of the neutralino $\mchi$ is given roughly by
$\mchi\simeq M_1\simeq 0.4\mhalf$. On the other hand, the mass of the lighter
stau $\staur$ (see Eq.~(\ref{sfmass:eq})) grows
somewhat more slowly with $\mhalf$, $\mchi\sim0.38\mhalf$. In the region of
large $\mhalf$ ($\gsim400\gev$) and small $\mzero$, $\staur$ (and in fact also
$\widetilde e_R$ and $\widetilde\mu_R$) become lighter than
the lightest neutralino. For a fixed $\mhalf$, as $\mzero$ grows, so does
$m_{\widetilde l_R}$ and $\chi$ becomes the LSP again.
Insisting on the neutralino LSP provides a very important constraint on the
plane ($\mhalf,\mzero$), excluding the region $\mhalf\gg\mzero$.
We note, however, that the regions where $\chi$ is not the LSP correspond to
large $\mgluino\gsim1\tev$. Also, we never
find the sneutrino to be the LSP: regions of small $\mhalf$ where this could
take place have been excluded by LEP.
We thus find that, in the most interesting region of low-energy SUSY it is the
neutralino which is most often the LSP. It is also mostly gaugino-type
(bino-type) -- this will be discussed in more detail
in Section~\ref{compassresults:sec}.

\subsubsection{Constraints from the Age of the Universe}\label{ageresults:sec}

For gaugino-like $\chi$'s the relic abundance $\abundchi$ depends most strongly
on the mass of the lightest exchanged sfermion in
$\chi\chi\ra f\bar f$; roughly
$\abundchi\propto m_{\tilde f}^4/\mchi^2$~\cite{mydmreview}. All sfermion
masses grow with increasing $\mzero$, and in the case of sleptons
much more slowly with $\mhalf$, so one expects that the bound $\abundchi\lsim
1$ which
results from requiring that the age of the Universe be at least 10 billion
years (see Section~\ref{age:sec}),
will be stronger for $\mzero$ than for $\mhalf$.
This is indeed often the case in the remaining regions of the parameter space.
The constraint~(\ref{tenbillionyrs:eq}) excludes large values of
$\mzero$ roughly above $1\tev$ and often even above a few hundred GeV.

For small $\tanb$ ($\tanb\approx 1$), the bound~(\ref{tenbillionyrs:eq}) is
typically much stronger and excludes $\mzero\gsim300\gev$ and
$\mhalf\gsim1\tev$. As $\tanb$ grows slightly to at least moderate values (2
and above), the bound becomes less constraining primarily
for $\azero$ around zero or positive allowing for somewhat larger values of
$\mzero$ and also opening
the region $\mzero\sim\mhalf$ above
1\tev.
This is because the $s$-channel $Z$-exchange in the process $\chi\chi\ra f\bar
f$ and the $\chi$ pair annihilation into pairs of of light Higgs bosons $h$
become unsuppressed and can reduce the LSP relic abundance. The $Z$ pole effect
is clearly visible
in the region $\mzero\gg\mhalf\simeq120\gev$ (see, \eg,
Figs.~\ref{envsone:fig}a, \ref{envsone:fig}d, or ~\ref{envstwo:fig}c).
But it is also in the region near this pole (and likewise near the $\hl$
pole) that the exact calculation of the relic abundance becomes difficult.
We have highlighted these regions in Fig.~\ref{casetwodetails:fig}.

The process $\chi\chi\ra h h$ is rarely dominant but it
can reduce the relic abundance considerably, especially in the most
interesting region of $\mhalf$ and $\mzero$ in the range of a few hundred~\gev\
for larger values of $\tanb$. This is clearly visible in
Fig.~\ref{casetwo:fig} (see also Fig.~\ref{veffs:fig}) where the
region to the right of an ``island'' of $\abundchi>1$ (large $\mzero$ and
$\mhalf\sim270\gev$) is again allowed because
the final state $hh$ becomes kinematically allowed. This effect is not present
for small $\tanb\approx1$ (compare Fig.~\ref{caseone:fig})  because the
coupling $h\chi\chi$ vanishes there.

Overall, the bound $\abundchi\lsim 1$ typically provides a very stringent
constraint on the regions of the parameter space not already
excluded by other criteria. It excludes $\mzero$ roughly above 1\tev,
except for large $\mhalf$ where some SUSY sparticle masses (\eg, $\mgluino$)
become very much larger than 1\tev\ and are therefore
disfavored by the fine-tuning criterion.

\subsubsection{Constraints from Requiring No Fine-tuning}
\label{ftresult:sub}

Finally, it is clear that if SUSY is to replace the SM as an effective theory
at the electroweak scale, its
mass parameters should not be much larger than $\mz$. Stated differently, since
the combination of $m_1^2$ and $m_2^2$ in
Eq.~(\ref{eq:halfmz}) has to give $\mz^2$, one would have to tune those
parameters to a high precision, unless they were broadly within a 1\tev\ mass
range~\cite{bgfinetuning}.
This fine-tuning in the potential minimization is a
remnant of the fine-tuning
exhibited by the full theory. In the full theory, one would parametrize
fine-tuning most naturally by $f\equiv\Lambda^2_{SUSY}/\mz^2$.
Instead, because radiative EWSB connects the SUSY scale to the electroweak
scale, we choose to parametrize it by
\be
f\equiv |m_1^2|/\mz^2
\label{ftuning:eq}
\ee
which is particularly
stable in terms of the running of the RGEs and the minimization of the
1-loop effective Higgs potential. (At the tree-level our definition is similar
but not identical to the definition of Ross and Roberts~\cite{rr}.)
The concept of fine-tuning is somewhat
subjective and various authors have used different definitions and criteria.

Fig.~\ref{ft:fig} shows
the typical scaling of the fine-tuning constant with the scale of SUSY for a
sample choice of input parameters.
In order to exclude regions where large fine-tuning must be invoked, we will
later place
an upper bound of $f\leq 50$.
As we can see from Fig.~\ref{ftscatter:fig}, this criterion typically
selects the heaviest sparticle masses below roughly 1\tev. It is worth
stressing however that, for large $\tanb$, both $\mgluino$ and $\msq$
can be significantly larger without any excessive fine-tuning. Thus
simple cuts $\mgluino,\msq<1\tev$ often made in the
literature~\cite{angut,klnpy1,klnpy2} may in
general be too strong.

One might hope that physics constraints would eliminate the need
for adding a separate fine-tuning constraint. That indeed is the case for large
ranges of parameters, which is very encouraging. For example,
for large $\mt=170\gev$
we find that the constraint $\abundchi<1$ cannot be
satisfied if $\mzero$ or $\mhalf$ are larger than several hundred~\gev.
This is also true for smaller $\mt$ if $\tanb$ is close to one.
In general much larger $\mzero$ and $\mhalf$ become allowed as $\tanb$ grows,
but this does demonstrate the kind of argument that
might lead to physical constraints on the parameter space in place of
fine-tuning~\cite{roberts}.

We will not apply the constraint $f\leq50$ in the rest of this section because
we
also want to display the
asymptotic behavior of solutions at very large values of $\mhalf$ and $\mzero$,
but will do so in Section~\ref{implications:sec} where we study the
implications
of this work for SUSY searches at accelerators.
We will see that, for some choices of $\mt$,
$\tanb$, and $\azero$, both $\mzero$ and $\mhalf$
are bounded from above by purely physical criteria,
and no fine-tuning constraint is needed.

\subsection{{\underline{CO}}nstrained {\underline{M}}inimal {\underline
{PA}}rameter{\underline{S}} {\underline{S}}pace (COMPASS)}
\label{compassresults:sec}

\subsubsection{General Properties}\label{genprops:sec}

We now focus on the region of the parameter space consistent with all the
constraints listed above. This region certainly meets our expectations for
where SUSY might be realized because the gauge couplings unify there,
correct EWSB takes place, and the experimental and cosmological constraints are
satisfied. In this constrained region of parameters (COMPASS) we now analyze
the various relations that result between the SUSY spectra and the implications
for SUSY searches.
Next, we will study
what additional restrictions are implied by imposing the dark matter
constraint.

Several typical examples of solutions resulting from our analysis
are presented in more detail in Figs.~\ref{caseone:fig}-\ref{casethree:fig} and
in Tables~\ref{caseone:table}-\ref{casethree:table}. In the graphs we
show the typical ranges of several interesting parameters.
In the tables we display the lowest and largest values of various masses
selected after scanning
all the choices of parameters compatible with COMPASS. (The ranges
selected by DM, presented in the last two columns, will be discussed shortly.)
We see that the allowed mass ranges are rather broad and typically allow for
masses as light as, or not much heavier than present experimental
limits.
\begin{table}
\centering
{\small
\begin{tabular}{|c||c|c||c|c||c|c|}
\hline
Mass Limits & \multicolumn{2}{c||} {COMPASS}  & \multicolumn{2}{c||}
{CDM} &
\multicolumn{2}{c|} {MDM} \\ \cline{2-7}
(GeV) & lower & upper & lower & upper & lower & upper  \\
\hline\hline
$\hl$ & 61 & 79 & 62 & 73 & 61 & 71 \\ \hline
$\ha$ & 635 & 1934 & 691 & 1340 & 658 & 1115 \\ \hline
$\widetilde e_L$ & 183 & 595 & 241 & 403 & 208 & 332 \\ \hline
$\widetilde e_R$ & 111 & 408 & 190 & 267 & 141 & 207  \\ \hline
$\widetilde\tau_1$ & 110 & 407 & 190 & 267 & 140 & 207 \\ \hline
$\widetilde\tau_2$ & 183 & 595 & 241 & 403 & 208 & 332 \\ \hline
$\widetilde\nu_L$ & 176 & 592 & 236 & 400 & 202 & 328 \\ \hline
$\widetilde u_L$ & 550 & 1621 & 571 & 1129 & 559 & 943 \\ \hline
$\widetilde u_R$ & 530 & 1549 & 552 & 1082 & 539 & 905 \\ \hline
$\widetilde t_1$ & 342 & 1199 & 354 & 810 & 347 & 660 \\ \hline
$\widetilde t_2$ & 607 & 1546 & 620 & 1112 & 612 & 948 \\ \hline
$\chi_1^0=$LSP       & 97 & 356 & 97 & 233 & 97 & 189  \\ \hline
$\neuttwo$ & 182 & 669 & 183 & 440 & 182 & 356	\\ \hline
$\charone$ & 180 & 668 & 182 & 440 & 180 & 355 \\ \hline
$\gluino$ & 596 & 1780 & 597 & 1234 & 598 & 1028\\ \hline
\end{tabular}}
\caption{ The lower and upper limits for the case
$\mt=145\gev$,
$\tanb=1.5$, $\azero/\mzero=0$, and $\sgnmu=-1$
(Fig.~\protect\ref{caseone:fig}) for all the solutions in COMPASS
(with no fine-tuning cut satisfying $f\leq50$ imposed), and for the subset of
solutions selected by
either the MDM
or CDM constraint. Because of the finite-size grid in our numerical sampling
the limits presented here could be somewhat relaxed and should
be treated only as indicative.
}
\label{caseone:table}
\end{table}
\begin{table}
\centering
{\small
\begin{tabular}{|c||c|c||c|c||c|c|}
\hline
Mass Limits & \multicolumn{2}{c||} {COMPASS}  & \multicolumn{2}{c||}
{CDM} &
\multicolumn{2}{c|} {MDM} \\ \cline{2-7}
(GeV) & lower & upper & lower & upper & lower & upper  \\
\hline\hline
$\hl$ & 91 & 113 & 96 & 112 & 94 & 108 \\ \hline
$\ha$ & 209 & 1773 & 346 & 1402 & 314 & 1115 \\ \hline
$\widetilde e_L$ & 118 & 1208 & 202 & 1047 & 175 & 1022 \\ \hline
$\widetilde e_R$ & 84 & 1070 & 162 & 1015 & 127 & 1007 \\ \hline
$\widetilde\tau_1$ & 81 & 1067 & 160 & 1012 & 125 & 1004 \\ \hline
$\widetilde\tau_2$ & 120 & 1206 & 203 & 1046 & 176 & 1021\\ \hline
$\widetilde\nu_L$ & 89 & 1205 & 187 & 1044 & 157 & 1019 \\ \hline
$\widetilde u_L$ & 326 & 2175 & 480 & 1965 & 417 & 1368 \\ \hline
$\widetilde u_R$ & 318 & 2095 & 465 & 1877 & 407 & 1310 \\ \hline
$\widetilde t_1$ & 196 & 1615 & 310 & 1552 & 250 & 1064 \\ \hline
$\widetilde t_2$ & 408 & 2011 & 538 & 1877 & 481 & 1333 \\ \hline
$\chi_1^0=$LSP     & 29 & 437 & 76 & 435 & 59 & 286 \\ \hline
$\neuttwo$ & 63 & 801 & 122 & 789 & 106 & 521 \\ \hline
$\charone$ & 51 & 800 & 112 & 788 & 101 & 520 \\ \hline
$\gluino$ & 294 & 2153 & 502 & 2151 & 419 & 1491 \\ \hline
\end{tabular}}
\caption{ The same as in Table~\protect\ref{caseone:table} but
for
$\mt=145\gev$, $\tanb=5$, $\azero/\mzero=-1$, and $\sgnmu=-1$
(Fig.~\protect\ref{casetwo:fig}).
}
\label{casetwo:table}
\end{table}
\begin{table}
\centering
{\small
\begin{tabular}{|c||c|c||c|c||c|c|}
\hline
Mass Limits & \multicolumn{2}{c||} {COMPASS}  & \multicolumn{2}{c||}
{CDM} &
\multicolumn{2}{c|} {MDM} \\ \cline{2-7}
(GeV) & lower & upper & lower & upper & lower & upper  \\
\hline\hline
$\hl$ & 113 & 131 & 116 & 125 & 114 & 119 \\ \hline
$\ha$ & 532 & 1502 & 564 & 1020 & 532 & 828 \\ \hline
$\widetilde e_L$ & 244 & 1069 & 244 & 1011 & 244 & 832 \\ \hline
$\widetilde e_R$ & 167 & 1023 & 167 & 1004 & 167 & 824 \\ \hline
$\widetilde\tau_1$ & 144 & 980 & 144 & 960 & 144 & 788 \\ \hline
$\widetilde\tau_2$ & 250 & 1051 & 250 & 991 & 250 & 816 \\ \hline
$\widetilde\nu_L$ & 230 & 1066 & 230 & 1008 & 230 & 828 \\ \hline
$\widetilde u_L$ & 641 & 1681 & 677 & 1156 & 641 & 931 \\ \hline
$\widetilde u_R$ & 631 & 1611 & 654 & 1110 & 631 & 924 \\ \hline
$\widetilde t_1$ & 441 & 1302 & 501 & 883 & 464 & 607 \\ \hline
$\widetilde t_2$ & 584 & 1579 & 687 & 1117 & 605 & 814 \\ \hline
$\chi_1^0=$LSP & 28 & 353 & 34 & 232 & 34 & 152 \\ \hline
$\neuttwo$ & 51 & 657 & 62 & 432 & 62 &	281 \\ \hline
$\charone$ & 50 & 657 & 61 & 432 & 61 &	281 \\ \hline
$\gluino$ & 207 & 1812 & 249 & 1257 & 249 & 874\\ \hline
\end{tabular}}
\caption{ The same as in Table~\protect\ref{caseone:table} but
for
$\mt=170\gev$, $\tanb=20$, $\azero/\mzero=0$, and $\sgnmu=-1$
(Fig.~\protect\ref{casethree:fig}).
}
\label{casethree:table}
\end{table}

On the other hand, we see
that, without constraining $\mhalf$ and $\mzero$ from above by the fine-tuning
constraint, all the masses can (for some $\mt$ and $\tanb$)
become very large, with the squark, gluino, and
heavy Higgs bosons ($H$, $A$, and $H^\pm$) typically being
the heaviest and the sleptons, charginos, and neutralinos being significantly
lighter, except for $\mzero$ large and $\mhalf
\sim O(\mz)$
where $\msl\approx\msq\gg\mgluino$.
Very large values of $\mzero\gg\mhalf$
and large values of $\mhalf\gg\mzero$ are typically disallowed by
$\abundchi<1$, charged LSP, color breaking (tachyonic $\stopone$), and no EWSB.
But for many choices of parameters, one can only exclude both large $\mhalf$
and $\mzero$ by imposing the fine-tuning constraint.
Thus we see that without further constraints or criteria, COMPASS still allows
for a wide range of SUSY masses, though these masses are correlated in very
specific ways.

An important quantity in the MSSM is the Higgs/higgsino mass parameter $\mu$.
In contrast with $\mhalf$, $\mzero$, and $\azero$, $\mu$ does not break SUSY
and therefore {\em a priori} it could be much larger than $\mz$. Similarly,
while supergravity
suggests a value of order $\mz$ for $\mzero$ and $\mhalf$, it generically does
not say
anything about the origin of $\muzero=\mu(\mgut)$. On the other hand,
phenomenologically, it would be very surprising if one of the defining
parameters of the MSSM were much larger than others.
In our analysis $\mu$ is determined by the other input
parameters and the adopted constraints.
We find $|\mu|$ broadly in the range of values spanned by either $\mhalf$ or
$\mzero$. Two typical patterns can be identified. In the cases when the
constraint from EWSB does not exclude the upper left-hand part of
the ($\mhalf,\mzero$) plane, we find $|\mu|\sim\mhalf$ for $\mhalf\gg\mzero$
and
$|\mu|\sim\mzero$ for $\mzero\gg\mhalf$.
Otherwise, $|\mu|\sim\mhalf$ for small $\mzero$ but slowly decreases
as $\mzero$ grows. Overall, the values of $\mu$ resulting from the analysis are
closely related to $\mhalf$ and $\mzero$ and only for such values of $\mu$ does
the CMSSM appear to be self-consistent.

One important consequence is that the lightest neutralino $\chi$ is in most
cases gaugino-like (more specifically, bino-like)~\cite{roberts}.
(Figs.~\ref{caseone:fig}-\ref{casethree:fig} and
Tables~\ref{caseone:table}-\ref{casethree:table} show typical neutralino mass
ranges and compositions; see also Fig.~\ref{apps:fig20}). This is quite a
remarkable theoretical prediction of the CMSSM in light of the fact that
a bino-like neutralino has been selected theoretically
as the unique attractive candidate for (neutralino) dark
matter~\cite{mydmreview,chiasdm}. Notice also that, while $\chi$ is typically
at least 80\% (and in most cases 90\%) bino, it is never a {\em pure} bino
state. Various analytic approximations for $\abundchi$ and related bounds
derived for a pure bino may thus be misleading~\cite{os}.

It is worth noting that
the LSP has typically a dominant bino component because $|\mu|$ almost always
comes out somewhat larger than $M_2\simeq0.8\mhalf$ (compare
Fig.~\ref{mum2:fig}). The composition and
scaling
properties of the neutralinos and charginos in the plane ($\mu,M_2$) have been
well-understood~\cite{chiasdm,ehnos,os}. In particular, for $|\mu|\gsim M_2$
the lightest neutralino is mostly gaugino-like (in fact, bino-like; see, \eg,
Fig.~1 and the discussion of gaugino purity in Ref.~\cite{chiasdm}). For
gaugino-like LSP the masses roughly satisfy the relations
\be
\mchi\simeq M_1\simeq 0.5M_2,\label{mchimone:eq}
\ee
\be
m_{\chi^0_2}\simeq m_{\chi^\pm_1}\simeq 2\mchi,\label{mchitwo:eq}
\ee
\be
m_{\chi^0_{3,4}}\simeq m_{\chi^\pm_2}\simeq |\mu|.\label{mchiheavy:eq}
\ee
(These approximations improve as $|\mu|\gg M_2$.)
It is important to note that in this approach, these relations
are characteristic to most solutions in COMPASS,
and do not come from GUT-dependent
constraints, such as proton decay.

In some regions of the ($\mhalf,\mzero$) plane we do find LSPs with significant
higgsino components. This happens for both $\mhalf$
and $\mzero$ small ($\lsim100\gev$), the region typically excluded by
experiment. It also happens in relatively small regions close to where EWSB
cannot be
achieved. There $|\mu|$ is smaller than $\mhalf$.
In a few other cases we also find higgsino-like LSPs for larger $\mhalf$ and
$\mzero$. This happens
for small $\mt$, $\tanb$ well above one, and very large $\azero$ (\eg, for
$\mt=120\gev$, $3\lsim\tanb\lsim20$, $\azero/\mzero=3$, $\sgnmu=\pm1$) in a
relatively limited region of large $\mhalf\approx \mzero\gsim400\gev$
disfavored by fine-tuning
(compare Fig.~\ref{ft:fig}). Higgsino-like LSPs
have been shown, however, to provide very little relic
abundance~\cite{mydmreview}. For
$\mchi>\mz,\mw,\mt$ the $\chi$ pair-annihilation into those respective
final states ($ZZ$, $WW$, $t\bar t$) is very strong~\cite{os}. Both below and
above those thresholds, there are additional co-annihilation~\cite{gs}
processes of the LSP with $\charone$ and $\chi^0_2$, which in this case are
almost
mass-degenerate with the LSP. Co-annihilation reduces $\abundchi$ below any
interesting level~\cite{dn,coann:japan}. Higgsino-like LSPs thus do not solve
the DM problem.
Except for those relatively rare cases,
we find an LSP of at least 80\% bino purity.

It is also interesting to explore what values of $\mu$ at the GUT scale
($\muzero$) result from the analysis. We choose to display it in terms of the
ratio $|\muzero|/\mzero$. A very simple relation emerges: $|\muzero|/\mzero$
decreases from a few in the large $\mhalf$ and small $\mzero$ region down to
one or less in the opposite extreme. If, for some choices of parameters, the
ratio falls down to zero, no proper EWSB occurs.

One other parameter of the model is $B$, which does not run very
much between its GUT value $\bzero$ and $B(\mz)$ (see Table~\ref{apps:tab2}).
A typical tendency
is for $B/\mzero$ to grow with $\mhalf$ and decrease with
$\mzero$. We also show in Fig.~\ref{bzeroazero:fig}
a scatter plot of $\bzero$ vs.~$\azero$ for $\mt=145\gev$ for all the
solutions belonging to COMPASS.
The value of $\bzero$ that we obtain as an output of our procedure rarely
yields
the relation $\bzero=\azero-\mzero$ that is often imposed by other analyses.

In this Section we have focussed mostly on the case $\mt=145\gev$. Varying
$\mt$ leads to significant modifications but the general features remain, as
can be seen by comparing Figs.~\ref{caseone:fig}-\ref{casethree:fig}.
We find for $\mt=170\gev$ that the constraint coming from imposing proper EWSB
becomes much weaker. Similarly, the regions where the lightest neutralino is
not the LSP become pushed towards even larger $\mhalf$. For a given point
in
the  ($\mhalf,\mzero$) plane $\alphas(\mz)$ grows with $\mt$ mostly due to the
$\sinsqthw$ dependence on $\mt$ (Eq.~(\ref{eq:s2w})), as discussed in
Section~\ref{running:sec}.

\subsubsection{Regions Favored by the Dark Matter
Constraint}\label{dmresults:sec}

We now point out the sub-region of COMPASS which is favored by the hypothesis
that the LSP is the dominant component of either cold or mixed dark matter. As
we discussed in
Section~\ref{dm:sec}, there is now abundant evidence for the existence
of DM in the Universe. The neutralino has become one of the most
attractive candidates for DM. In the pure CDM scenario one expects the
LSP relic abundance to be in the range given approximately
by~(\ref{abundrangecdm}), while in the currently more favored mixed (CDM+HDM)
scenario it should roughly satisfy the range~(\ref{abundrangemdm}).

Applying either~(\ref{abundrangecdm}) or~(\ref{abundrangemdm}) to the parameter
space under consideration results in selecting only relatively narrow bands in
the plane ($\mhalf, \mzero$) whose shape
and location vary with other parameters but typically correspond to
both $\mhalf$ and $\mzero$ in the range of a few hundred GeV. (See
Figs.~\ref{caseone:fig}-\ref{casethree:fig}.) Of course, they fall into the
region constrained by the age of the Universe ($\abundchi<1$).

More importantly, requiring enough DM (\ie, taking lower limits in
either~(\ref{abundrangecdm}) or~(\ref{abundrangemdm})) typically leads to {\em
lower} limits on both $\mhalf$ and $\mzero$ and, as a result, also on the SUSY
mass spectra which are higher than in COMPASS alone.
It is interesting that the mass ranges consistent with
either~(\ref{abundrangecdm}) or~(\ref{abundrangemdm}) are typically less
accessible at LEP~II and FNAL.
This can be seen by comparing the lower limits allowed by COMPASS with those
selected by the CDM or MDM scenarios in
Tables~\ref{caseone:table}--\ref{casethree:table}. (See also
Table~\ref{apps:tab2} and the discussion in Section~\ref{dmaspects:sec}.) For
example, in the case presented in Fig.~\ref{casetwo:fig} and
Table~\ref{casetwo:table} applying the DM
constraints causes the chargino $\charone$ and the sleptons to be completely
inaccessible to LEP~II and the gluino to be above the reach of FNAL. It also
makes it harder to discover $\hl$ and other particles.
On the other hand, the DM constraint severely lowers the upper ranges of
masses for all the particles making them much more likely to be accessible
at future accelerators like the NLC or LHC. Prospects of searches
for various particles will be discussed in more detail in
Section~\ref{implications:sec}, and in particular
the detectability of the lightest Higgs as a function of LEP~II beam energy
will be analyzed in Section~\ref{detection:sec}. Here we only note that,
with large enough $\sqrt{s}$, $\hl$ has a very
good chance of being discovered at LEP~II.

While one might argue that the constraints~(\ref{abundrangecdm})
or~(\ref{abundrangemdm}) do not carry the same weight as some other constraints
listed above, they do reflect our current cosmological
expectations and serve as a strong guide to those regions
of the parameter space in which SUSY solves the DM problem.

\subsection{Effect of the Full Effective Higgs Potential}
\label{veffresults:sec}

The results from our analysis have also served to reinforce the need for
using the full 1-loop effective potential in the minimization
procedure~\cite{grz}.
We already argued in Section~\ref{formal:sec} that the 1-loop contributions
to $\vhiggs$ were important in order to stabilize the scale-dependence of the
potential, but one can also see the net effect of using the full 1-loop
$\vhiggs$ in our model-building results. As well, one can see the smaller role
played by
the non-leading contributions to $\vhiggs$, that is, contributions not coming
from the $t-\stopq$ splitting~\cite{klnpy2,florida2,anveff,casas}.
In Fig.~\ref{veffs:fig}, we
have shown two plots of the ($\mhalf,\mzero$) plane for the
choice of input parameters as in Fig.~\ref{casetwo:fig}.
Fig.~\ref{veffs:fig}a shows the
region of parameter space allowed after we have excluded the regions in
which EWSB did not occur (labelled E), where the LSP was charged or
colored (labelled L), and where the neutralino relic abundance would
``overclose''  the
Universe (labelled A), for the renormalization group-improved tree
level potential only. On the other hand, Fig.~\ref{veffs:fig}b shows the
parameter
space available for the same choices of parameters, but now with the
renormalization group-improved 1-loop effective potential with leading terms
only. Notice that the regions in which EWSB did not occur have enlarged, taking
over some of the regions which were excluded before on the basis of their LSP
being electrically charged.
However, the strong bound placed on the parameter space by DM constraints has
considerably weakened, leaving the region in which $\mzero\simeq\mhalf
\rightarrow$~large available, pending a fine-tuning cut. We note also that the
effect of including the 1-loop contributions to $\vhiggs$ is negligible in the
region $\mhalf\ll\mzero$ favored by
the proton decay constraint~\cite{angut,ln:proton} in the minimal $SU(5)$
GUT.
Finally, including all (leading and non-leading) terms does not modify
the situation sizably as can be seen by comparing Figs.~\ref{veffs:fig}b
and~\ref{casetwo:fig}a.
The qualitative difference is extremely small, which
we found to be a general result.


\section{Applications}
\label{implications:sec}

\subsection{Overview}\label{overview:sec}

The analysis described in previous sections has led us to a restricted
parameter space for $\mt$, $\tanb$, $\mhalf$, $\mzero$, $\azero$, and
$\sgnmu=\pm1$ in the CMSSM which we call COMPASS (see
Section~\ref{compassresults:sec}).  Most previous studies of SUSY predictions
have
preferred to fix some of these parameters by assumptions and vary one
or two, either with or without constraints.  This is useful and interesting
and can lead to instructive predictions, but there is always doubt about
their generality.

We have taken the alternative approach of studying the fully constrained
parameter space described in Section~\ref{results:sec}.
We know that any point in COMPASS is already
guaranteed to have gauge coupling unification, a Higgs mechanism, all
phenomenological constraints satisfied, \etc\  We can then ask a variety
of questions about the regularities of the resulting solutions, whether they
have predictions of interest, and so forth. For example, we can ask:  what
fraction
of solutions gives a spectrum of sparticles that can be detected at LEP~II and
FNAL (or any other present or future facility), and in what channels
do we most expect to find sparticles?  What do the
solutions predict for BR($\bsg$), $\Gamma(Z\to b\bar b)$?  What is
$\abundchi$ for the solutions?  When new experimental or
theoretical information is available it can be easily added to constrain
the parameters further.
In the following we describe
a number of such results.  More specifically, in this section
we examine the solutions that pass all the theoretical, experimental, and
cosmological constraints listed in
Sections~\ref{constraints:sec}--\ref{results:sec}, \ie, solutions in COMPASS.
We impose two additional cuts. We keep only those solutions which require no
large
fine-tuning of parameters. We take $f\leq50$ which roughly corresponds to the
heaviest squark, gluino, and Higgs masses falling below 1\tev, except for very
large
$\tanb$ where $\mgluino$ and $\msq$ can be larger
(see Section~\ref{ftresult:sub}).
We also impose
the lower bound BR($b\to s\gamma )>1.5\times 10^{-4}$~\cite{cleo7} (see
Section~\ref{brbsg:sec}). The solutions in this restricted set will be called
``acceptable.'' Most of the results presented in this section
have been derived with $\mt=145\gev$, but in some cases we consider
other values of $\mt$.

Recall that our constraints do not require a detailed knowledge of the
physics at the high scale.  Our parameter space is intended to
be the most general one which is independent
of multifarious GUT scenarios.  It is for this reason that
we do not impose a constraint on the lifetime
of the proton.
The proton decay constraints
have been included first by Arnowitt and Nath~\cite{angut}, and also by
Lopez,~\etal~\cite{ln:proton} mainly in an $SU(5)$ GUT.  They find a longer
proton lifetime for smaller tan${\beta}$ and smaller $\mhalf$, so
this region of the parameter space is enhanced for them.  We will study
the implications of adding assumptions about unification and a
GUT group in the near future.

We have also assumed a common scalar
mass $\mzero$ and a common gaugino mass $\mhalf$, both of which can
be relaxed, which we will consider in the near future.  Keeping these comments
in mind, we now discuss some CMSSM (Constrained MSSM) results.

\subsection{Higgs Physics}\label{higgsphysics:sec}

\subsubsection{What is $\mhl$ due to?}\label{whatismh:sec}

In a supersymmetric theory with electroweak symmetry breaking the Higgs
boson mass is calculable, and it is very interesting to ask what
parameters in th theory play a role in determining the value of
$\mhl$.  The tree level mass matrix for the two CP--even scalar bosons is
\be
M^2=\pmatrix{-B\mu\,\tanb+{1\over 2} (g_1^2+g_2^2) v^2
\cos^2\beta &
B\mu -{1\over 2} (g_1^2+g_2^2) v^2 \sin\beta\,\cos\beta \cr
B\mu -{1\over 2} (g_1^2+g_2^2) v^2 \sin\beta\,\cos\beta &
-B\mu\,\cot\beta+{1\over 2} (g_1^2+g_2^2) v^2
\sin^2\beta \cr}
\ee
where $\mz^2={1 \over 2} (g_1^2+g_2^2) v^2$ and $v^2=\vone^2+\vtwo^2$.
If $B\mu =0$ or if $v^2=0$ then this matrix has a zero eigenvalue.
Thus in supersymmetry one cannot think of $\mhl$ as coming only from the Higgs
self--interaction.
Any interpretation is complicated since $B\mu$, $\tanb$, and $v^2$ are
all involved.  Furthermore, the one-loop effective potential can yield $\mhl$
significantly above the tree-level result~\cite{radcorrs}.

Haber~\cite{haber8} has emphasized that there is a lower limit on $\mhl$;
even if the tree
level value is zero the one--loop potential generates a mass.  He finds
a lower value above 60\gev, but that assumes 1\tev\ squark masses.
We agree that there is a lower limit, but it is sensitive to squark masses, as
shown in Fig.~\ref{apps:fig1}
where we plot the {\em lower limit} of $\mhl$ versus
$\sqrt{\mstopone\mstoptwo}$ which contributes the largest radiative correction
to $\mhl$.  The lower limit basically arises
because of the way the EW breaking comes about in SUSY.  Eq.~(\ref{eq:s2beta})
leads to a lower limit on $B\mu (=m_3^2)$ and thus a lower limit
on $\mh$.

Effectively, $\mhl$ arises from three sources:  the product of
SUSY parameters $B\mu$, the value of the vevs (whose sum in quadrature
is fixed numerically by $\mz$), and the one-loop radiative corrections.
To demonstrate how these sources of Higgs mass interplay, we show
in Figs.~\ref{apps:fig2a} and \ref{apps:fig2b}
plots of $\mhl$ vs.~$\tanb$
and $\mhl$ vs.~$\sqrt{|B\mu|}$, where $\mhl$ is the full radiatively
corrected Higgs mass.  The upper limit on $\mh$ is due to the usual
argument that in SUSY the Higgs self--coupling is fixed by the
gauge couplings with an additional contribution from the radiative
corrections~\cite{haberedkane}.

One can see a strong correlation between
$\tanb$ and the allowed $\mhl$.  This is expected since
the tree level upper bound for $\mhl$ goes like $| {\rm cos} 2\beta |$.
In fact, we find from our solutions that, for $\mt=145\gev$, $\tanb<5$
if
$\mhl<85\gev$.  Therefore, if LEP~II finds the Higgs then $\tanb$
is constrained to be less than 5 for all our surviving solutions.  Solutions
with $\tanb\geq 5$ and $\mhl<85\gev$ are excluded mainly by
one of three effects:  (1) the chargino or sneutrino mass
is too low; (2)
the LSP is not the neutralino; or, (3) electroweak symmetry breaking
does not occur. Fig.~\ref{apps:fig11} shows the distribution of $\mh$ for all
acceptable models with $\mt=145\gev$.

\subsubsection{Detection of the Higgs Boson}\label{detection:sec}

Interestingly, we find essentially no solutions for which
the $\hl+\ha$ mode is detectable
at $\sqrt{s}\lsim$ 210\gev\ (fewer than 0.1\% of the solutions),
since $\mha$ is too large for all CMSSM.

Almost all solutions have $\sin^2(\beta-\alpha)>0.98$, so the $Z\hl$ cross
section (which is proportional to sin${}^2(\beta-\alpha)$) is not
suppressed~\cite{sinalpha}. Thus the experimental limit on the SM Higgs boson
effectively applies to the $\hl$ of the MSSM in all acceptable solutions.
The current LEP bound is $\mhl\gsim 62\gev$~\cite{higgs7}.  That
sin${}^2(\beta-\alpha)\approx 1$ and that $\mha$ is not small enough for
the $\hl+\ha$ channel to be accessible at LEP are related.

A similar result holds for the $t\bar t h$ coupling.  It has a factor
cos$\alpha$/sin${\beta}$ which is within a few percent of one over
essentially the entire set of solutions,  so that methods to detect $\hl$
by radiation off a top quark will work essentially as well for the
SUSY $\hl$ as for the SM Higgs.

In Fig.~\ref{apps:fig3} we show the percent of solutions with $\mt=145\gev$
for which $\hl$ is detectable at a given $\sqrt s$.
One can see that about 30\% of
the solutions are detectable when LEP energy increases to 178\gev, increasing
to
about 75\%
if $\sqrt s$ is increased up to 210\gev;  for the MSSM 100\% is reached
at $\sqrt s\simeq 220\gev$.  This limit is well known since in the MSSM
the upper limit on $\mhl$ is about 125\gev\ for $\mt\lsim 150\gev$.

A number of groups~\cite{kunszt7}
have examined the detectability of at least one SUSY
Higgs at LEP or SSC/LHC.  They concluded that much of the complete
parameter space could be covered, but not all.  Results were often
presented on a $\tanb$ vs.~$m_A$ plot.  In Fig.~\ref{apps:fig4} we show
where our constrained solutions appear on such a plot.
We also mark
the approximate region within which the detection of at
least one SUSY Higgs was found unlikely~\cite{kunszt7}.
Amusingly,
about 2/3 of the solutions that do fall in the region are
detectable at LEP~II or FNAL in some other channel.

We also present in Fig.~\ref{apps:mhma} a scatter plot of $\mhl$ vs.
$\mha$ for all acceptable solutions for $\mt=145\gev$. The distinct branches
seen in the graphs correspond to different choices of $\tanb$. One can see how
$\mhl$ grows with  $\mha$ and
$\tanb$. The dependence is smeared to some extent
by varying all the other parameters of the model.

\subsection{What Is the Origin of $\mt$?}\label{origin:sec}

We have remarked on several aspects of the role of $\mt$ in other
sections.  Here we discuss briefly the questions of what contributes
to the mass of $\mt$.  Of course, the main question is why
$\mt \gg \mb$ and how that is answered in supersymmetry.
More explicitly, once we are below the scale where EW breaking
has occurred, one can write at a scale Q,
\be
\mt(Q)=h_t(Q)v(Q)\tanb(Q)/\sqrt{1+{\rm tan}^2\beta(Q)},
\ee
for the top quark running mass, and likewise for the bottom,
\be
\mb(Q)=h_b(Q)v(Q)/\sqrt{1+{\rm tan}^2\beta (Q)}.
\ee
If the $SU(2)$ symmetry
were not broken here, we might expect $\mt$ to be only a little larger than
$\mb$.  To understand the effects that can enter, we can take the ratio
at $\mt$:
\be
{\mt(Q=\mt)\over \mb(Q=\mt)}={h_t(\mt)\over h_b(\mt)}\tanb
\ee
and we can define
\be
\label{apps:req}
r\equiv{h_t(\mt)/h_b(\mt)\over h_{t0}/h_{b0}}
\ee
where $h_{t,b0}\equiv h_{t,b}(\mx)$. Then finally
\be
\frac{\mt(\mt)}{\mb(\mt)}=r \frac{h_{t0}}{h_{b0}} \tanb.
\ee
Thus the large ratio $\mt(\mt)/\mb(\mt)\approx 50$ could be
due to any of three factors:  $\tanb$, the ratio of the Yukawa couplings at the
high scale, and/or the RGE running of the Yukawa couplings, as expressed by
the value of $r$.

The RGEs of the top and bottom Yukawa couplings are
\be
{dh_t \over d\ln Q}={h_t\over 8\pi^2}\left( -{8\over 3} g_3^2-{3\over 2}g_2^2
-{13\over 30} g_1^2+3h_t^2+{1\over 2} h_b^2 \right) ,
\ee
\be
{dh_b \over d\ln Q}={h_b\over 8\pi^2}\left( -{8\over 3} g_3^2-{3\over 2}g_2^2
-{7\over 30} g_1^2+3h_b^2+{1\over 2} h_t^2+{1\over 2} h_\tau^2 \right) .
\ee
Since the $g_1^2$ and $h_\tau^2$ contributions are numerically very small,
we see that if $h_b\approx h_t$ at the high scale, they will run
down together,
in which case the large value of $\mt$ is generated predominantly by
large $\vtwo$, and necessarily tan$\beta\simeq \mt/\mb$ is large.  For $h_t$
larger than $h_b$ at the high scale, $\mt$ increases relative to
$\mb$ from the running, and the physical $\mt$ is reached with a smaller
$\tanb$. To illustrate these effects we show in Fig.~\ref{apps:fig5}
a scatter plot of $h_{t0}/h_{b0}$
vs.~$\tanb$ for the constrained solutions, all with
$\mt=145\gev$.
We see that $(h_{t0}/h_{b0})\tanb\simeq 60$ is a good approximation
to the results except for very large and small $\tanb$, so that
the large value of $\mt$ cannot be interpreted as coming from the
running; it must be input, either as a large GUT-scale ratio
$h_{t0}/h_{b0}$ or as a large $\tanb$.  In particular, if $\hl$ is discovered
at
LEP178, then the large $\mt$ must be due to the top Yukawa at the
GUT scale (compare Fig.~\ref{apps:fig2a}).

\subsection{BR($b\to s\gamma$)}\label{bsg:sec}

The recently reported upper bound BR($b\to s\gamma)<5.4\times 10^{-4}$
(see Section~\ref{brbsg:sec})
has spurred an increased interest in predictions for $b\to s\gamma$
in SUSY.
Barbieri and Giudice~\cite{barbieri7}
have reminded us that in the limit of unbroken
supersymmetry the MSSM prediction (including the Standard Model part)
is zero due to a theorem of Ferrara and Remiddi~\cite{ferrara7},
and they have
shown that SUSY solutions will give reasonable values for this rate.
Garisto and Ng~\cite{garisto7}, and others, have also done a general SUSY
analysis of the implications of BR($b\to s\gamma)$.

In Fig.~\ref{apps:fig6} we show a histogram of
BR$(b\to s\gamma)$ for all the solutions.  We have checked that the solutions
in the peak come from all over the parameter space and in no sense represents
a decoupling region. For a typical solution the magnitudes of the $W$-$t$ loop,
the $\hcpm$-$t$ loop, and the $\chi^\pm$-$\stp$ loop contributions are all
about the same, with the $W$-$t$ and $\hcpm$-$t$ loops having the same sign and
the $\chi^\pm$-$\stp$ loop having the opposite sign.
We see that the CMSSM naturally produces solutions in
the right range. These results are predictions in the sense that
here we impose
no constraint on the model space from $b\to s\gamma$ data. (For other
uses of the model space we cut at the upper and lower limits indicated in
the figure, so that our model space does include the BR($b\to s\gamma$)
constraint in general, except for the present discussion.)

Some authors~\cite{hewett7}
have claimed that this decay strongly constrains charged
Higgs boson masses.  To show that there is no strong constraint in the CMSSM,
we show in Fig.~\ref{apps:fig7} a plot of BR$(b\to s\gamma)$ vs.~$m_{H^+}$ for
the solutions in the region between the upper and lower limits. That is,
every $m_{H^+}$ in Fig.~\ref{apps:fig7} gives a BR($b\to s\gamma$) consistent
with experiment.

Finally, it is interesting to look at BR$(b\to s\gamma)$ vs.~$\tanb$
in Fig.~\ref{apps:fig8}.
Smaller $\tanb$ values concentrate somewhat in the allowed region,
though acceptable solutions occur at any $\tanb$.  In the large
$\tanb$ region
the chargino contribution can be quite large and
negative~\cite{oshima7,garisto7}.

\subsection{Detection of SUSY at LEP~II and FNAL}\label{lepfnal:sec}

There are a number of possible ways to detect supersymmetric partners
at FNAL or LEP~II.
We estimate that about 32\% of all CMSSM acceptable solutions (see
Sec.~\ref{overview:sec})
have either a superpartner or the light Higgs boson detectable
at LEP with $\sqrt s=178\gev$ and 500 pb${}^{-1}$, or at
FNAL (with, say, 500 pb${}^{-1}$ integrated luminosity),
or both ($\hl$ will only be detectable at LEP~II, not FNAL).
We include in this sample all
solutions with $\mt=145\gev$, $\mzero\leq 1\tev$, $\mhalf\leq 1\tev$, and
$f\leq50$;
that is conservative, giving $\widetilde{q}$ and $\gluino$ masses over 2\tev\
for the largest $\mzero$, $\mhalf$.  The solutions with large $\mzero$,
$\mhalf$
are generally not accessible at FNAL or LEP~II, but they also do not
increase the number of acceptable solutions rapidly enough to dilute the
32\% result even if larger $\mzero$, $\mhalf$ were to be included.
The results do not vary rapidly with $\mt$.  These numbers are for
solutions with $\abundchi<1$.

At LEP~II 24\% of the acceptable solutions allow detection via one-sided
events, $e^+e^-\to \chi^0_2 \chi^0_1$, where $\chi^0_1\equiv\chi$, followed by
$\chi^0_2\to l^+l^-\chi^0_1$.  And
18\% will have pair production
of the lightest chargino, and 8.6\% detection of $\hl$.  There is overlap,
of course, and 30\% of all solutions are detectable at LEP~II.  Selectrons
are detectable in 3.6\%, a light stop 0.7\%, and $\hl+\ha$ in 0.06\%
of the solutions.

An interesting way---perhaps the only way at LEP---to determine if $\hl$
is a SUSY
Higgs is to measure the cross section for $e^+e^-\to \hl+{\rm
nothing}$~\cite{hkkq}.
In the Standard Model this entire cross section should be from
$e^+e^-\to \hl(\to b\bar b)+Z(\to \nu\bar \nu)$ and will be very accurately
known.  In SUSY there is also a contribution from
$e^+e^-\to \chi^0_2(\to \hl+ \chi^0_1)+ \chi^0_1$; unfortunately
this contribution
is larger than 10\% of the SM cross section in only 0.6\%
of the solutions at LEP~178 but would be the ``proof'' of SUSY in these cases;
at larger $\sqrt{s}$ the fraction of solutions where this effect could be
observed increases rapidly.

As an illustration, we show in Fig.~\ref{apps:fig23} what fractions of the
$(\mzero,\mhalf)$ plane would be constrained
by SUSY searches at LEP~II if $\mt=145\gev$. The
kinematic criteria that we use to determine the detectability at LEP178 are
$\mcharone<85\gev$, $\msf<85\gev$ (for any sfermion),
$\mneutone+\mneuttwo<170\gev$,
$\mhl+\mha<170\gev$, and $\mhl+\mz<170\gev$. We also require that any
event-signature lepton have energy above $5\gev$ or any quark have energy above
$10\gev$.
The regions marked by crosses (empty boxes)
will always (never) be accessible to LEP~II for any combination of input
parameters. Filled boxes mark the regions accessible for some
combinations of parameters.
In window (a) we show the combination
of possible SUSY searches at LEP~II by applying the criteria listed above. In
window (b) we show the same for the chargino $\charone$ alone, and
in window (c) for the lightest Higgs assuming $\mhl<80\gev$.
($\hl$ is a very SM-like Higgs.) Finally,
window (d) shows how much larger a region would be explored by searching
for $\hl$ up to 110\gev. Remember that very large values of $\mhalf$
are disfavored by the fine-tuning constraint (compare, \eg,
Fig.~\ref{ftscatter:fig}).

At FNAL gluino detection will occur in 11\% of all solutions, squark detection
in 5\%, detection of $\chi^0_1 \chi^{\pm}_1$ in 25\%,
$\chi^{\pm}_1\chi^{\mp}_1$ in 14\%, $\chi^0_2\chi^0_1$ in 24\%,
$\chi^0_2\chi^{\pm}_1$ in 12\%, $\widetilde{t}_1+\bar{\widetilde{t}_1}$ in 4\%.
These combine to make
26\% of all solutions being detectable at FNAL.  The FNAL/LEP
overlap is large, so combining them only increases the percentage of
solutions detectable at FNAL or LEP~II to the
above mentioned 32\%.

The kinematic criteria that we use to determine the detectability at FNAL are
$\mhcpm<\mt-5\gev$, $\mgluino<300\gev$, $\mcharone<85\gev$,
$\msq<300\gev$, and
$\mneutone+\mneuttwo<170\gev$. We also require that any event-signature
lepton or quark have energy above $15\gev$.
Some of the percentages listed above for the detectability of different
channels will decrease, particularly at FNAL, when detection efficiencies
and cuts to reduce background are considered.  But with sufficient
luminosity and sufficiently good detectors the above numbers should be
approached.  We are presently undertaking full simulations
of signals and backgrounds to determine reliable signatures and strategies.
There have been previous studies of FNAL and LEP~II detectability
in some depth~\cite{det7}.  Our only advance so far
over some of these is that their conclusions are
based on a parameter space some
parts of which are excluded because the various constraints are not
satisfied.

We note that at FNAL, for gluinos lighter than 300\gev, 80\% of solutions
have squarks heavier than gluinos, so that the appropriate way to
simulate $\gluino$ detection is to take $m_{\widetilde{q}}>m_{\gluino}$
in the first approximation.

We have also investigated our CMSSM parameter space to see how
the top quark search at FNAL could be affected by supersymmetry.
We find that for $\mt=170\gev$ approximately 3\% of all acceptable
solutions kinematically allow one or more of the following:
$t\to b H^{\pm}$ decay,
$t\to \widetilde{t}_1 \chi^0_1$ decay, or
$\gluino\to \widetilde{t}_1 t$
with $\gluino<250\gev$. (For smaller $\mt$ this fraction is reduced.)
Any one of these kinematic possibilities can significantly alter the
kinematic analysis and/or the effective rates (after cuts)
of top quark production.  Once the top physics at FNAL settles into place
if none of these is observed then the parameter space is reduced
a few percent.

Some reduction in detectable solutions would occur if the
$\chi^{\pm}_1-\chi^0_1$ mass difference were so small that the resulting
lepton or jet from $\chi^{\pm}_1$ decay were too soft to detect.
Fig.~\ref{apps:fig9} shows a plot of
$m_{\chi^{\pm}_1}-m_{\chi^0_1}$ vs.~$m_{\chi^{\pm}_1}$ from which we see that
most solutions have no problem here; and Fig.~\ref{apps:fig10} shows the energy
of
the lepton or jet from $\chi^{\pm}_1$ decay.

We will report a study on how effective higher energy linear colliders
will be at studying SUSY for the constrained model space later.  For
now we note that NLC with $\sqrt s=350\gev$ will be able to detect
$\hl$ and at least one superpartner for about 75\% of the constrained
solutions;  this number grows to about 97\% as $\sqrt s$ grows to 500\gev.

If $\hl$ is not detected at LEP~II once $\sqrt s\approx$ 210\gev, most
but far from all solutions
will be excluded, particularly if $\mt\approx$ 145\gev.  Fig.~\ref{apps:fig11}
shows a histogram of $\mhl$ values and one
can see that most solutions are below
110\gev.
In solutions with a Higgs sector extended beyond that of the minimal
one there is still an upper limit on $\mhl$, but it can be as large
as about 146\gev~\cite{kane7}.

Overall, we conclude that, while not finding superpartners
at LEP~II and FNAL does eliminate nearly a third of the
parameter space, it will still leave many possibilities open.

\subsection{What If FNAL and LEP~II Do Not Detect a
Sparticle?}\label{no_det:sec}

If superpartners are not detected at LEP~II or FNAL then much of the
low ($\mzero$, $\mhalf$)
parameter space can be excluded.
Bounds on $\mhalf$ are determined mainly by the bounds on the
gluino.  The gluino mass is related to $\mhalf$ by
\be
\mhalf=\xi_{\gluino} m_{\gluino}.
\ee
We find the lower limit of
$\xi_{\gluino}=\alphax/\alpha_s(m_{\gluino})$ to be (compare text below
Eq.~(\ref{mtwomgluino:eq}))
\be
\xi_{\gluino}>0.36.
\ee
So if $m_{\gluino}$
is determined from experiment at FNAL to be
greater than 300\gev\ then
\be
\mhalf>(0.36) (300\gev)=107\gev.
\ee
Bounds
can also be placed on $\mhalf$ by direct searches on the lightest
chargino.  By taking the square ($MM^T$) of the chargino mass matrix, we
find that
\be
m_{\chi^{\pm}_1}^2<\xi_{\chi^{\pm}_1} \mhalf^2+2 m_W^2 {\rm cos}^2\beta.
\ee
where
\be
\xi_{\chi^{\pm}_1}\equiv {M_2^2 \over \mhalf^2}<{16 \over 25}.
\ee
So if $m_{\chi^{\pm}_1}>M_W$ then the $\tanb$--dependent
bound for $\mhalf$ is
\be
m_{1/ 2}>m_W {5 \over 4} \sqrt{1-2{\rm cos}^2\beta}
\ee
For high $\tanb$ this bound ($\sim 100\gev$)
would become comparable to the resulting bound on $\mhalf$ from
the gluino search.
In any case, bounds
on $\mhalf$ are obtained straightforwardly from direct sparticle
searches.

It is more difficult to put bounds on $\mzero$.  Although
the squarks and sleptons gain mass from $\mzero$ they also have
contributions from the gaugino masses (compare Eq.~(\ref{sfmass:eq})).
Therefore bounds on $\mzero$ from direct
searches of squarks and sleptons must be presented as a function
of $\mhalf$.  Letting $\widetilde{f}$ represent any squark or slepton
and $\hat m_{\widetilde{f}}$ the lower mass bound of $\widetilde{f}$
we can write the inequality relation
that constrains the ($\mzero,\mhalf$) plane:
\be
\label{ellipse_eq}
\mzero^2+b_{\widetilde{f}_{L,R}} m^2_{1/2}>\hat
m^2_{\widetilde{f}}-m_f^2 \mp \mz^2\cos2\beta \left[T_3^{f_{L,R}}
-Q_{f_{L,R}}\sinsqthw\right].
\ee
For example, for $\widetilde{f}=\widetilde{e_L}$,
\be
{9\over 20}<b_{\widetilde{e_L}}<{1\over 2}.
\ee
The upper and lower bounds on $b_{\widetilde{e_L}}$ are good to within
5\% accuracy.
As can be clearly seen from Eq.~(\ref{ellipse_eq})
the slepton bound translates to an
ellipse (first quadrant) in the ($\mzero$, $\mhalf$) plane where all
($\mzero,\mhalf$) inside the ellipse are excluded.

\subsection{What If FNAL or LEP~II Does Find a Sparticle?}\label{do_det:sec}

If FNAL or LEP~II discover one or more sparticles we would like to extract
from this the GUT-scale Lagrangian.  That is, we would like to extract
the supersymmetric input parameters
$\mzero$, $\mhalf$, $\azero$, $\tanb$ and $\sgnmu$ from all the
observables that are sensitive to them.  (We assume
that we know $\mt$).  In this section, we briefly discuss how this
should be done.  It turns out that this is more difficult than it
first appears; analytic methods are of limited applicability.

The equations used in the last section to put bounds on $\mzero$ and $\mhalf$
given bounds on the gluino and sfermions can also be used to pin down
$\mzero$ and $\mhalf$.  In general, each fixed value of an observable,
whether it be a mass or a cross section or an asymmetry or anything
else, generates a hypersurface in input parameter space ($\mzero$, $\mhalf$,
$\tanb$, $\azero$, $\sgnmu$). (The gluino
mass, for example, would predominantly determine $\mhalf$ up to additional
uncertainties discussed below.) The
effective dimensionality of this surface is determined by how many
input parameters significantly affect the value of the observable.
Determining this hypersurface is very difficult because
of all the nonlinearities in relating low energy parameters to
the input parameters
through self--consistent solutions of many
RGEs.  In the last section and in
Refs.~\cite{martin7} there are some simple equations that allow us to
estimate functional relations between input parameters and sparticle
masses.  Even though these simple relations are often a good
approximation to the full analysis
we must keep in mind that every observable necessarily
depends on all input parameters, though with varying importance.  For example,
a precise
determination of $\mhalf$ from $\mgluino$ is limited because
$\alpha_s(m_{\gluino})$ depends in general on all sparticle thresholds
(at 1--loop)
\be
m_{\gluino}=\frac{\alpha_s(m_{\gluino})}{\alphax}
\mhalf=f(\mzero,\mhalf,\azero,\tanb,\sgnmu).
\ee
Likewise, precise gaugino contributions to sfermion masses require detailed
knowledge of the sparticle thresholds which introduce all input
parameters into the final determination of the sfermion masses.
Methods based on analytic expressions are clearly limited. They
may suffice to provide rough estimates of input parameters,
especially soon after the discovery of a given particle, when the experimental
bounds will be still large, but they cannot be improved to
accommodate a complex analysis based on several well-measured observables.

Our approach to the problem of determining input parameters from
low--energy observables does allow for such improvements. We explore wide
ranges of the input parameter space and let the computer do the
work.  We have really already employed this technique to generate
the COMPASS of the CMSSM.  We have cut hypersurfaces in input parameter space
with Higgs and sparticle mass
bounds, BR($b\to s\gamma$) limits, $\abundchi$ bounds, fine-tuning limit,
$\alpha_s(\mz)$, \etc\  However, highly precise measurements of observables
especially sensitive to supersymmetric loop corrections, or direct
measurements of sparticle production, will pin down the input parameters
which determine those low energy observables.

We present in Table~\ref{apps:tab3} an example of how several experimental
measurements can shrink the input parameters space to a ``point'' which would
in turn
allow us to predict all other observables (other sparticle masses,
cross sections, \etc).
\begin{table}
\centering
{\small
\begin{tabular}{||c|c|c|c|c|c||}
\hline
Detected Particle(s) & $\mzero$         & $\mhalf$    & $\azero /\mzero$ &
$\tanb$      & $\mu$   \\ \hline
$\hl$ & $\leq671$ & $\leq549$ & unbounded   & unbounded  & -472 -- 253 \\
\hline
$\charone$ & unbounded  & 74 -- 450    & unbounded & $\leq40$ & -477 -- 332 \\
\hline
$\stopone$ & $\leq549$ & 61 -- 202    & unbounded & $\leq20$ & -428 -- 420 \\
\hline
$\gluino$ & unbounded & 74  & unbounded & $\geq3$
      & -460 -- 205 \\ \hline
$\eL$ & $\leq74$ & 61 -- 111   & unbounded & $\leq5$     & -237 -- 172 \\
\hline
$\hl$, $\charone$ & $\leq368$ & 74 -- 136    & unbounded & $\leq3$ & -414 --
169 \\ \hline
$\hl$, $\charone$, $\stopone$ & $\leq302$ & 74 -- 111   & -3.0 -- -1.0  &
$\leq3$ & -414 -- 140 \\ \hline
$\hl$, $\charone$, $\stopone$, $\gluino$ & $\leq136$ & 74 & -3.0 -- -1.5
& 3 & 120 -- 140  \\ \hline
$\hl$, $\charone$, $\stopone$, $\gluino$, $\eL$ & 61 & 74 & -2.5 --
-2.0 & 3 & 120 -- 130 \\ \hline
\end{tabular}}
\caption{ Table demonstrating how input parameter space
($\mzero$, $\mhalf$, $\azero$, $\tanb$,
and $\sgnmu$) is constrained by detection of particles. The initial ranges of
input parameters are listed in Sec.~\protect\ref{effectsresults:sec}.
For this example we assume detection of these
particles to mean $\mh=80\pm 5\gev$, $m_{\charone}=70\pm 5\gev$,
$m_{\stopone}=110\pm 10\gev$,
$m_{\gluino}=210\pm 10\gev$ and $m_{\eL}=95\pm 5\gev$.  If more
than one particle type is listed in the ``Detected Particle(s)''
column then the range for each of the input parameters is
found from knowledge of {\em all listed} particle masses.
Keep in mind that the ranges of parameters listed in the table are values
obtained on our numerical sampling grid and therefore have errors associated
with them corresponding to the grid spacing. For example, when we quote
$\mhalf=74\gev$ we really mean that we find no acceptable solutions with
$\mhalf\leq61\gev$ (the next lowest $\mhalf$ value on our grid) and no
acceptable solutions with $\mhalf\geq91\gev$ (the next highest $\mhalf$ on our
grid). If our grid were very fine-grained, then we could quote
ranges that would more accurately reflect how well the parameters were
determined, and that reflect the experimental errors better.}
\label{apps:tab3}
\end{table}
The first row in Table~\ref{apps:tab3} lists
the range allowed for each input parameter given detection of just the
lightest Higgs ($\mh$ determined from experiment to be, say,
$\mh=80\pm 5\gev$).
The value of $\mu$ is evaluated at $\mz$ and its sign is the one of $\sgnmu$.
We see that just knowing the mass of the lightest Higgs
alone clearly does not significantly constrain any of the input
parameters.  The next row in Table~\ref{apps:tab3} lists the range allowed
for each input parameter given detection of just the lightest chargino
($\charone$ determined by experiment to be, say,
$\mcharone=70\pm 5\gev$).
Here again knowledge of just one mass, the lightest chargino, does not
significantly constrain input parameter space.  In the next three rows we
list the range of input parameters given detection of just $\mstopone$
($\mstopone=110\pm 10\gev$), just $\gluino$ ($\mgluino=210\pm 10\gev$),
and just $\eL$ ($m_{\eL}=95\pm 5\gev$).  The next row assumes
detection of $\hl$ {\em and} $\charone$.  Notice how the combination of
these two masses restricts $\mzero$, $\mhalf$ and $\tanb$
far beyond what knowledge of each mass can do individually.

As we progress down the rows of the table with each subsequent row
assuming more and more detected particles, the input parameters become
more and more constrained until $\mzero$, $\mhalf$, $\tanb$, and
$\sgnmu$ are determined precisely at the level of our numerical sampling.
Only $\azero$ remains stubbornly undetermined though it is better constrained.
This is a general rule about $\azero$:  few observables are very
sensitive to it.  Observables which depend on third generation sparticle
left--right mixing are most sensitive to $\azero$.

As this example shows,
by generating many self--consistent solutions and ``filling
up'' input parameter space with them, we have the means by which to
use all observables simultaneously to constrain input parameter space.
This approach of generating solutions with
the most precision possible, calculating observables without untrustworthy
simplifying assumptions, and then simultaneously comparing all the generated
solutions with all the calculated observables is a powerful way to analyze and
constrain minimal supersymmetry. It is this approach that will quickly
enable us to add better measurements of observables and any new
observables to the CMSSM, including possible announcements of
sparticle detection.  With our method we can go directly from data to
the parameters of the effective Lagrangian at the unification scale.

\subsection{The $\Gamma(Z\to b\bar b)$ Partial Width}\label{zbb:sec}

Precision measurements at LEP~II currently show a slight
deviation from the Standard Model value for
the extracted value of $\Gamma_{b\bar b}/\Gamma_{\rm had}$.
The current LEP average~\cite{lepreview} is
\be
R_b={\Gamma_{b\bar b}\over \Gamma_{\rm had}}=0.2200\pm 0.0027.
\ee
Several groups~\cite{sm_zbb7}
have studied the loop corrections to this partial width
in the Standard Model.  The SM prediction
for $R_b$ is heavily dependent upon the value of $\mt$, but~\cite{lepreview}
the predicted value is $1.5\sigma$ lower than
the measured value for $\mt=145\gev$ and it gets even lower for
higher $\mt$.

The supersymmetric contributions to this decay width have been calculated
in Refs.~\cite{djouadi7,boulware7}.  We use the equations of
Ref.~\cite{boulware7}
to calculate the supersymmetric contributions to $R_b$ within
the CMSSM.  We perform scalar integral reductions and numerical
calculations from Ref.~\cite{stuart7}.

In Fig.~\ref{apps:fig22} we plot the histogram of all acceptable
solutions with $\mt=145\gev$ and $0.16<\abundchi <0.33$.
Notice that within the CMSSM the supersymmetric contributions
tend to increase $R_b$.
This increase in $R_b$ is mainly due to light $\widetilde{t}_i$
and $\chi^{\pm}_i$.  Interestingly, if we require the
calculated $R_b$ to be within
$1\sigma$ of the measured value, then approximately
75\% of the resulting solutions will be detectable at LEP~II
and 83\% of the solutions will be detectable
at FNAL.  The channel which is by far most
detectable at LEP~II for these solutions is
$\chi^{\pm}_1\chi^{\mp}_1$ production.  At FNAL many
possible channels for detection of sparticles are allowed:
$\gluino$, $\chi_1^{\pm}\chi_1^{\mp}$,
$\chi^{\pm}_1\chi^0_1$,
$\chi^{\pm}_1\chi^0_2$,
and $\chi^0_1\chi^0_2$.

\subsection{Neutralino LSP as Dark Matter}\label{dmaspects:sec}

As we have already seen in Section~\ref{results:sec}, the relic abundance
$\abundchi$ of the lightest neutralino is an important quantity
which plays a significant role in constraining the parameter space
of the CMSSM. If the Universe is at least 10 billion years
old then $\abundchi<1$ (and the Universe's age of
15 billion years or more gives $\abundchi\lsim0.25$),
see Section~\ref{age:sec}. We have also seen in
Section~\ref{compassresults:sec}
that over significant regions of the model parameter space the LSP
provides enough dark matter in either the CDM
($0.25\lsim\abundchi\lsim0.5$, see~(\ref{abundrangecdm}))
or currently more favored MDM
($0.16\lsim\abundchi\lsim0.33$, see~(\ref{abundrangemdm})) scenarios
(see Sec.~\ref{dm:sec}).
This is a remarkable property of the CMSSM given the fact that,
unlike in the case of
many phenomenological quantities, calculating $\abundchi$ involves
elements of the
physics of the early Universe and {\em a priori} the resulting
predictions for the LSP relic abundance could be completely incompatible
with the expectation of low-energy SUSY.

In this section we provide some more insights into the cosmological
properties of the neutralino LSP.
First, for
the restricted set of ``acceptable solutions", as described in
Sec.~\ref{overview:sec}, and for $\mt=145\gev$
we show in Fig.~\ref{mum2:fig} a scatter plot in the plane $(\mu,M_2)$. Notice
a large concentration of solutions below the diagonals $M_2=|\mu|$
corresponding to the LSP being mostly gaugino-like (see
also the discussion above and below Eq.~\ref{mchimone:eq}).
This property is even more explicitly pronounced in Fig.~\ref{apps:fig20} where
we show for $\mt=145\gev$
and $\tanb=10$ a scatter plot of the LSP
wavefunction ($Z_{11}^2$ for the bino, \etc)
with an additional constraint $0.16<\abundchi<0.33$ (MDM)
imposed.
Notice that the LSP is mainly a bino, as
expected~\cite{chiasdm}. This has been already illustrated for a few specific
cases in
Figs.~\ref{caseone:fig}-\ref{casethree:fig}.
A very similar plot can be made for any value of $\tanb$ and
changes little for the CDM scenario. Without the MDM (or CDM) constraint
imposed, we find in Fig.~\ref{apps:fig20} also some points with somewhat
smaller bino purity, with the largest concentration however still remaining at
large $Z^2_{11}$.

We also examine the predictions for the LSP relic density resulting
from our restricted set of acceptable solutions defined in
Sec.~\ref{overview:sec}, corresponding roughly to our expectations
for low-energy SUSY.
In Fig.~\ref{apps:fig21} we plot a histogram of $\abundchi$
for all otherwise acceptable solutions with $\mt=145\gev$.
Notice that there
is a strong peak at $\abundchi\approx 0.1$
suggesting that in the CMSSM the MDM
scenario is somewhat more favored relative to the CDM one.
Within the
framework of CMSSM we can view this result one of two ways, either
as a prediction for $\Omega_\chi$ (given $h_0$) for solutions
which satisfy all other criteria, or
that cold dark matter puts a severe constraint
on the CMSSM if we demand that the LSP contributes
most of the (cold) dark matter needed in the either the MDM or CDM scenarios.
Both viewpoints are quite constraining: the first viewpoint
for LSP (cold) dark matter and the second for the parameter space of
the CMSSM.

Finally, in Fig.~\ref{apps:relic:fig} we plot $\abundchi$ vs.~$\mhalf$,
$\mzero$, $\tanb$, and $\mu$.
(In these graphs we lift the constraint $\abundchi<1$.)
Notice that the ranges of $\abundchi$ favored by both MDM and CDM generally
select both $\mhalf$, $\mzero$, and $|\mu|$ (which is an output parameter in
our
analysis) in a broad region of a few
hundred~\gev. For small $\mhalf,\mzero\lsim100\gev$ the relic density is too
small because some sleptons are rather light there (roughly less
than 100\gev) which enhances the $t$-channel pair-annihilation $\chi\chi\ra
f\bar f$~\cite{chiasdm}.
Also notice that large $\tanb$ produces more solutions with
low $\abundchi$ than do the solutions with low $\tanb$ which is due to
a very strong enhancement in the LSP
pair-annihilation cross section caused by the exchanged
pseudoscalar $\ha$ ~\cite{sugradm}. (The coupling $\ha\chi\chi$
scales like $\tanb$.)

\subsection{The SUSY Spectrum}
\label{spectrum:sec}

By varying our input parameters over wide ranges of values, we can consider
what ranges of sparticle and Higgs masses one should expect from the CMSSM.
In Fig.~\ref{apps:fig13} we present a scatter plot of mass vs.
particle type for all the acceptable solutions in CMSSM with
$\mt=145\gev$. Little changes significantly with varying $\mt$.
In the next section, we will further prejudice our
parameter space in order to find sample solutions that are preferred
theoretically.

\subsection[Physics Prejudice Enhancement of Part of Model Space?]{Physics
Prejudice Enhancement of Part of Model \\ Space?}
\label{fit:sec}

In this section we apply some
experimental and theoretical prejudices to the acceptable solutions.
For example, solutions with
BR($b\to s\gamma)\approx 5.4\times 10^{-4}$ are less favored than
solutions with BR($b\to s\gamma)\approx 3.5\times 10^{-4}$.Furthermore,
theoretically we prefer solutions with lower fine-tuning.  These
are just two examples.  Other prejudices that we can apply on the
solutions are $\mb(\mx)/\mtau(\mx)\simeq 1$, the LSP providing the
right amount of cold dark matter, and $\alpha_s(\mz)$ close to
its experimental central value.  As can be gleaned from previous
sections, some of these prejudices work against others in
some respects.

We have attempted to select a subset of all the solutions which are
most likely to satisfy all (or most) of the above prejudices.  We do
that by effectively ``squeezing'' the solutions:  select a preferred
value for each constraint above and reducing the errors (or allowed
region) by a factor of 2.  In analyzing this squeezed set we find
that the fraction of solutions which are detectable at LEP~II and
FNAL goes down by about a factor of 2.
That is, only about 18\% of this set of solutions is detectable at
FNAL or LEP~II, whereas in the full set the fraction is 32\%.
We therefore
find that {\em the set of solutions which best satisfies our current
experimental and
theoretical prejudices are characteristically more difficult to
detect than the full set of solutions allowed by current experiment.}
Table~\ref{apps:tab2} presents three examples of such solutions.
Solutions 1 and 3 are not detectable at LEP~II or FNAL, but Solution 2 is in
the chargino, $\hl$ (LEP~II) and gluino (FNAL) channels.
We view this section as an initial attempt to add weighted physics
criteria in order to select a part of the model space to use for other
considerations such as phenomenological predictions or theoretical
studies.
\begin{table}
\centering
{\small
\begin{tabular}{||l|l|l|l||}
\hline
Model Parameters         & Solution 1  &  Solution 2   & Solution 3\\
\hline\hline
$\tanb$, $m_t$
&10, 145 & 1.5, 145 & 5, 170 \\ \hline
$\mzero$, $\mhalf$, $\azero/\mzero$
& 247, 302, -2.5 & 91, 111, 2.5 & 111, 247, 2.5  \\ \hline
$\bzero/\mzero$, $B(\mz)$
&-0.67, -0.03 & 2.84, 2.20 & 0.02, -0.46 \\ \hline
$\muzero$, $\mu (\mz)$, $\alpha_s(\mz)$
& 394, 450, 0.124 & -214, -218, 0.127 & 303, 304, 0.129 \\ \hline
$\alphax$, $\mx/10^{16}\gev$    &
0.041, 1.64   & 0.042, 2.21  & 0.041, 2.04  \\ \hline
$\hl$, $\hh$, $\ha$, $\hcpm$    & 116, 346, 345, 354   & 62, 317,
305, 315  &
113, 326, 324, 333  \\ \hline
$\widetilde{e}_L$, $\widetilde{\mu}_L$, $\widetilde{\tau}_L$  & 328,
328, 324
& 124, 124, 124 & 211, 211, 211         \\ \hline
$\widetilde{e}_R$, $\widetilde{\mu}_R$, $\widetilde{\tau}_R$  & 276,
276, 268
& 104, 104, 104 & 152, 152, 152             \\ \hline
$\widetilde{\nu}_{e_L}$, $\widetilde{\nu}_{\mu_L}$,
$\widetilde{\nu}_{\tau_L}$
& 318, 318, 318 & 114, 114, 114 & 197, 197, 197 \\ \hline
$\widetilde{u}_L$, $\widetilde{c}_L$, $\widetilde{t}_L$       & 700,
700, 634
& 283, 283, 299 & 570, 570, 555      \\ \hline
$\widetilde{u}_R$, $\widetilde{c}_R$, $\widetilde{t}_R$       & 677,
677, 620
& 276, 276, 271 & 551, 551, 492      \\ \hline
$\widetilde{d}_L$, $\widetilde{s}_L$, $\widetilde{b}_L$       & 705,
705, 622
& 288, 288, 266 & 575, 575, 534      \\ \hline
$\widetilde{d}_R$, $\widetilde{s}_R$, $\widetilde{b}_R$       & 676,
676, 667
& 276, 276, 276 & 550, 550, 549       \\ \hline
$\widetilde{\tau_1}$, $\widetilde{\tau_2}$             & 266, 326   &
104, 124
 & 151, 212   \\ \hline
$\widetilde{t_1}$, $\widetilde{t_2}$                   & 419, 705   &
177, 363
 & 445, 594          \\ \hline
$\widetilde{b_1}$, $\widetilde{b_2}$                   & 620, 670   &
264, 278
 & 533, 550          \\ \hline
$\chi^{\pm}_1$, $\chi^{\pm}_2$ 		  & 239, -468   &  53, 257 &
192, -329
\\ \hline
$\chi^0_1=$LSP, $\chi^0_2$, $\chi^0_3,\chi^0_4$   & 126, 239, -457, 464
& 27, 65,
-220, 263 & 102, 191, -316, 324      \\ \hline
$M_1$, $M_2$, $M_3=\gluino$             & 126, 245, 718   & 45, 90,
292 & 102,
200, 610      \\ \hline
BR$(b\to s\gamma)$            & $3.52\times 10^{-4}$ & $2.97\times
10^{-4}$ &
$4.76\times 10^{-4}$ \\ \hline
$m_b(\mx)/\mtau(\mx)$ & 0.794 & 0.750 & 0.794 \\ \hline
$\abundchi$            & 0.27 & 0.24 & 0.22\\ \hline
$\widetilde{B}$ \& gaugino purities & 0.99, 0.99 & 0.67, 0.87 &
0.98, 0.98
\\ \hline
\end{tabular}}
\caption{ Three representative solutions--one with
rather
light sparticles,
and the other two with intermediate to heavy sparticles.
All masses are given in GeV.
Some neutralino and chargino masses are
quoted as negative.  This is merely an indication of the phase of the
mass eigenstates (expressed as $\eta_i$ by some authors); we include
it in case people wish to use these number in calculations, but only
magnitudes should be considered as experimentally relevant.
$\widetilde{t}_{1,2}$ are physical eigenstates, while
$\widetilde{t}_{L,R}$ correspond to the stop mass-matrix entries.
Same for the stau and the sbottom.}
\label{apps:tab2}
\end{table}


\section{Summary and Comments}\label{summary:sec}

Encouraged by gauge coupling (grand) unification as implied by LEP,
we have made an attempt to frame SUSY by re-considering Minimal
Supersymmetry in the light of GUTs.
We have parametrized the whole
multitude of low-energy SUSY masses and couplings
in terms of just five free parameters (and the sign of $\mu$), including the
mass of the top quark $\mt$ which will soon be known.
We have demanded gauge coupling unification and proper
electroweak symmetry breaking. In further
reducing the allowed parameter space, we have included
all the relevant experimental and cosmological constraints that can be imposed
without choosing a specific GUT gauge group at the unification scale.
We have not found the present experimental bounds on SUSY particle
masses to be particularly constraining; in fact they are only
beginning to limit the lower range of the SUSY masses. In contrast,
rather robust cosmological constraints, like requiring that the
Universe be at least 10 billion years old and that the LSP not
be electrically charged, rule out large fractions of the SUSY
parameter space. Furthermore, much more specific
conclusions about the resulting SUSY mass spectra and properties can
be drawn
if one expects the neutralino LSP to be the
dominant dark matter component (in either cold- or mixed-DM scenarios)
in the flat Universe.

A number of groups have already reported studies along the same lines we
follow. Our work is more comprehensive and complete in that more of the
theoretical and phenomenological constraints are included than in any
previous work, and precision to the few percent level is required wherever
appropriate in a fully consistent manner.
We also include more applications than have been considered previously.

It is only by combining all the
constraints and exploring wide ranges of parameters that
one is able to establish where SUSY might be realized. Remarkably,
we find that SUSY is preferably realized in the
range of Higgs, sfermion, and gaugino masses of several hundred GeV and
below,
with larger values sometimes allowed by our constraints but disfavored by too
much
fine-tuning. At this point one still cannot favor any range of $\mt$, unless
one insists on the $\mb-\mtau$ unification which
in most cases (but not always) implies a very heavy top quark. Similarly, all
values of $\tanb$ between one and about 50-60 (perturbative upper bound) are
still
allowed, although the resulting phenomenology often differs considerably in the
small and large $\tanb$ regime. On the other hand, significant constraints can
be placed on the ($\mhalf,\mzero$) plane. The region $\mhalf\gg\mzero$
is invariably excluded by requiring either electroweak symmetry breaking or
neutral LSP, while $\mzero\gg\mhalf$ is typically ruled out by either  EWSB or
a lower bound on the age of the Universe. Refs.~\cite{angut,ln:proton}
have argued that the region
$\mzero\gg\mhalf\approx\mz$ (and small
$\tanb\lsim8$) appears to be favored by bounds on
the proton decay in the simplest SUSY
$SU(5)$ model, but we prefer not to rely on an $SU(5)$ GUT, so this
region is not favored for us.

We have made a first survey of phenomenological
implications for future SUSY searches in high-energy experiments. We
find reasonable chances for eventually finding a chargino at LEP~II and the
gluino or gauginos at the Tevatron. The light Higgs $\hl$ has a very good
chance of
being discovered at LEP~II but most likely only if its
beam energy is pushed close to or beyond 200\gev. On the other hand, the
chances are very slim with the currently approved $\sqrt{s}=178\gev$. The LHC
will produce large quantities of all superpartners.

Several predictions follow from our analysis which could have
served to falsify the CMSSM before the sweeping supercollider
searches for the squarks, Higgs, and the gluino are done.
\begin{itemize}
\item
We derive a general {\em upper} bound on $\alphas(\mz)<0.133$.
For larger $\mt$, and for some regions of $\tanb$ when $\mt$ is smaller,
there is a lower bound $\alphas(\mz)\gsim0.117$. In the regions where
the bound is not implied by the physics constraints, it is implied instead
by the addition of a fine-tuning constraint.
GUT-scale threshold
corrections may be sizeable and modify these limits by several percent.
(In this paper we have ignored all
GUT-scale corrections because we have not yet studied
specific unification gauge groups and their Higgs structures, although we do
assume $\sinsqthw(\mgut)=3/8$.)
\item
Within our parameter space the light Higgs mass is very SM-like and
$\mhl\lsim120\gev$ for $\mt=145\gev$ and $\mhl\lsim130\gev$ for $\mt=170\gev$,
with somewhat lower values usually favored.
We also find that $\mhl>85\gev$ for $\tanb>5$.
If $\hl$ had been discovered below about $30\gev$, our entire parameter space
would have been excluded.
\item
The charged Higgs is always significantly heavier than $\mw$ and
its discovery should not be expected at LEP~II and most likely even
at the NLC500. Other heavy Higgs bosons ($\hh$ and $\ha$) are
almost degenerate in mass with $\hcpm$.
If $\hcpm$ is discovered below about $110\gev$, our entire parameter space is
excluded.
\item
If BR$(Z\to b\bar b)<0.214$ and $\mt\lsim150\gev$, all solutions are excluded.
Similar bounds exist for larger $\mt$.
\item
The LSP is almost invariably
of the gaugino-type (more precisely bino-type), as advocated early in
Ref.~\cite{chiasdm}. If the cold dark matter is not of this type, almost
all solutions from this study are exluded. If at least one sfermion
(other than the stop or sneutrino) had always
been lighter than about 80\gev, there
would have been too little neutralino DM~\cite{chiatlep2}.
Furthermore, had the sneutrino been the LSP, the CMSSM would not have predicted
enough DM.
\item
Several clear patterns and relations among the masses of the Higgs and SUSY
particles arise which can be tested in future accelerators.
\end{itemize}

We have also addressed several related issues recently discussed in the
literature. We agree with~\cite{ekn,anselmo,rr,lp1} about
the need to  use two-loop RGEs and to
include multiple thresholds in considering the running of the gauge
couplings. One-loop RGEs and simplified one-step approaches
can each lead to
errors in $\alphas(\mz)$ of some 10\% (while the experimental error
is at the 5\% level).
We emphasize
that it is inappropriate to use the so-called
effective SUSY-breaking scale
$\msusyeff$ in deriving the GUT mass scale $\mgut$ and gauge coupling
$\alphax$. In particular, the value of $\mgut$ derived this way can be twice
that coming from the full two-loop calculation.
We have noted that GUT-scale corrections
of 10-15\% to the relation $\mb(\mgut)=\mtau(\mgut)$ may have a
significant impact on the resulting mass of the bottom
quark $\mb(\mb)$ and may allow for this relation to hold for wide
ranges of both $\mt$ and $\tanb$.
We also find it very important to use the one-loop effective Higgs
potential particularly in the range of large masses ($\gsim1\tev$)
where it can lead to qualitatively different conclusions than had we
simply used the tree level potential.

Remarkably, we find that for larger $\mt$ it is possible to place upper
bounds on all superpartner masses of $O(1\tev)$, without imposing any
fine-tuning criterion. For smaller $\mt$ this is still true for some regions of
$\tanb$ near one, but not for all.

Ultimately the goal is to go from experimental data to a determination of the
effective Lagrangian of the supersymmetric and (perhaps) unified theory at a
scale
of $O(10^{16}\gev)$. We have shown by example that an effective and perhaps
optimal
way to do this as data becomes available is to systematically reduce the
allowed
parameter space numerically. Once the high-scale Lagrangian is known, perhaps
the
patterns among its parameters will lead toward an understanding of how SUSY is
broken
and what the underlying theory is.

Minimal Supersymmetry is a very attractive theoretical framework which
makes several falsifiable phenomenological and cosmological predictions while
at the same time encompassing all the remarkable experimental
successes of the Standard Model. The most natural ranges for supersymmetric
particle masses typically lie above the reach of currently operating
accelerators (LEP, FNAL) but may be accessible to
the upgraded LEP and FNAL, and should be successfully
explored by the next generation of colliders.

\section*{Acknowledgments}

This work was supported in part by the US Department
of Energy. We have benefited from conversations with R.~Arnowitt,
M.~Einhorn, H.~Haber, S.~Kelley, B.~Lynn, M.~Peskin, N.~Polonsky,
P.~Ramond, R.~Roberts, J.F.~Zhou
and many others.



\begin{figure}[tbhp]
\caption{ Regions in the $\mt-\tan\beta$ plane
consistent with bottom-tau Yukawa unification. The region bounded
by the solid lines represents the region of parameter space consistent
with $\mb/\mtau=1$ at $\mgut$
for $4.7\leq\mbpole\leq 5.1\gev$. The region between
the dashed lines is consistent with $\mb/\mtau=0.9$ at $\mgut$.
Here we have taken the effective scale of SUSY to be $90\gev$ and
$\alphas(\mz)=0.120$.}
\label{mbone:fig}
\end{figure}

\begin{figure}[tbhp]
\caption{ Same as Fig.~\protect\ref{mbone:fig} but now with
$\alphas(\mz)=0.112$.
Notice that the available parameter space has increased markedly.}
\label{mbtwo:fig}
\end{figure}

\begin{figure}[tbhp]
\caption{The running of the sparticle masses from the GUT scale to the
electroweak
scale, for a sample set of input parameters (see ``Solution 3'' in
Table~\protect\ref{apps:tab2}
later in this paper). The bold lines are the three soft gaugino masses,
$m_\gluino$, $M_2$ (labelled $\widetilde W$) and $M_1$
(labelled $\widetilde B$). The light
solid lines are the squark ($\squark_L$, $\squark_R$, $\stopl$,
$\stopr$) and slepton ($\slepton_L$, $\slepton_R$) soft masses, where we
ignore $D$-term contributions and the mixing of the stops for this figure.
Finally, the dashed  lines represent the soft Higgs masses, $m_1$ and $m_2$
(see Eq.~\protect\ref{vtree:eq}),
labelled by $H_d$ and $H_u$. The onset of EWSB is signalled by $m_2^2$ going
negative, which is shown on the plot as $m_2$ going negative for convenience.
}
\label{running:fig}
\end{figure}

\begin{figure}[tbhp]
\caption{ Plots of the ($\mhalf,\mzero$) plane showing regions
excluded by lack of EWSB (labeled E), neutralino not being the LSP (L), the age
of the Universe less than 10 billion years (A), $\mcharone<47\gev$ (C),
BR$(b\to s\gamma)>5.4\times10^{-4}$ (B), and SM-like lightest Higgs mass
$\mhl<60\gev$ (H).
We take $\mt=145\gev$, $\sgnmu=-1$, and several representative choices of
$\tanb$
and $\azero$. In window
(a) $\tanb=1.5$, $\azero/\mzero=0$,
in (b) $\tanb=5$, $\azero/\mzero=0$,
in (c) $\tanb=5$, $\azero/\mzero=-2$, and
in (d) $\tanb=20$, $\azero/\mzero=3$.
In window (a), the regions excluded by each criterion are identified
separately, while for windows (b)--(d) only the total envelope is shown.
For each case, the limit imposed by our fine-tuning constraint $f\leq50$
is shown
as a dotted line, disfavoring regions above and to the right of the line.
Notice the importance of combining several different criteria in constraining
the parameter space. (Only the most limiting constraints are marked.) Note that
in window (a) the ($\mhalf,\mzero$) allowed region is bounded entirely by the
physics constraints, without a fine-tuning constraint, though $\mhalf$
extends  to larger values than allowed by this constraint (see
also Fig.~\protect\ref{caseone:fig}).
}
\label{envsone:fig}
\end{figure}

\begin{figure}[tbhp]
\caption{ Same as in Fig.~\protect\ref{envsone:fig} but for
$\mt=145\gev$, $\sgnmu=+1$, and in
(a) $\tanb=5$, $\azero/\mzero=0$,
in (b) $\tanb=5$, $\azero/\mzero=-2$,
in (c) $\tanb=5$, $\azero/\mzero=2$, and
in (d) $\tanb=10$, $\azero/\mzero=-2$.
}
\label{envstwo:fig}
\end{figure}

\begin{figure}[tbhp]
\caption{ Same as in Fig.~\protect\ref{envsone:fig} but now with
$\mt=170\gev$, $\sgnmu=-1$ and
(a) $\tanb=5$, $\azero/\mzero=0$,
in (b) $\tanb=5$, $\azero/\mzero=-2$, and
in (c) $\tanb=20$, $\azero/\mzero=3$.
In (d) we take $\tanb=10$, $\azero/\mzero=-2$, and $\sgnmu=+1$.
Generally,
for $\mt=170\gev$ we find both $\mhalf$ and $\mzero$ bounded from
above by physical constraints.
}
\label{envsthree:fig}
\end{figure}

\begin{figure}[tbhp]
\caption{The ($\mhalf,\mzero$) plane for $\mt=145\gev$, $\tanb=1.5$,
$\azero/\mzero=0$, and $\sgnmu=-1$. We show in window
(a) the region allowed by all constraints (dark solid)
(compare Fig.~\protect\ref{envsone:fig}a). We also mark the bands favored by
the CDM (between dashed lines) and MDM scenarios (between dotted lines)
and the limit imposed by our fine-tuning constraint $f\leq50$ (light solid). We
also plot in window (b) $\alphas(\mz)$ at 0.120 through 0.132 in steps of
0.002, decreasing for larger $(\mhalf,\mzero)$ with 0.120 solid,
0.126 dashed and
any others dotted;
in (c) $\mcharone$ (solid), $m_{\widetilde l_L}$ (dashes), and $m_{\widetilde
l_R}$ (dots) at 45, 80, 150, and 250\gev. (In the two latter cases only the
last three values occur in the graph.) In window (d)
we plot $\mchi$ (solid) at 18 (thick), 45, 75, 100, 125, and 150\gev. We also
display the gaugino purity ($Z^2_{11}+Z^2_{12}$, dots) of 0.8 and 0.9
increasing to the right.
In window (e) we plot $\mgluino$ (dots) and average $m_\squark$ (other than
$\mstopq$) (solid) at 250, 500, 750, and
1000\gev, and in window (f) $\mhl$ between 60 and 130\gev\ in 5\gev\ intervals
(60, 120 dark solid, 90 dark dashes, all others light
solid). Mass contours in each window increase with increasing
$(\mhalf,\mzero)$.
}
\label{caseone:fig}
\end{figure}

\begin{figure}[tbhp]
\caption{ Same as in Fig.~\protect\ref{caseone:fig} but for
$\mt=145\gev$, $\tanb=5$, $\azero/\mzero=-1$, and $\sgnmu=-1$.
Window (a) also demonstrates the importance of including final states other
than
$f\bar{f}$ (in this case $hh$) in the calculation of $\abundchi$
(see Section~\protect\ref{ageresults:sec}).
}
\label{casetwo:fig}
\end{figure}

\begin{figure}[tbhp]
\caption{ Same as in Fig.~\protect\ref{caseone:fig} but for
$\mt=170\gev$, $\tanb=20$, $\azero/\mzero=1$, and $\sgnmu=-1$.
}
\label{casethree:fig}
\end{figure}

\begin{figure}[tbhp]
\caption{ Region in the ($\mhalf,\mzero$) plane ruled out by the LEP
bound on the
sneutrino mass, $m_\sneutrino>43\gev$, for both small and large $\tanb$ and by
the CDF bound $\mgluino>141\gev$ (dashes, cascade decays neglected).
The chargino mass bound $\mcharone>47\gev$ often (but not always) leads to
additional excluded regions.}
\label{sneutrino:fig}
\end{figure}

\begin{figure}[tbhp]
\caption{
For the case presented in Fig.~\protect\ref{casetwo:fig}    ($\mt=145\gev$,
$\tanb=5$, $\azero/\mzero=-1$, and $\sgnmu=-1$) we delineate the regions close
to the $Z$ (window (a)) and $\hl$ (window (b)) poles in the process
$\chi\chi\ra\bar f f$
where our calculation of $\abundchi$ cannot be trusted.
(The effect of other poles is much less significant.)
In window
(a) $|\mchi-\mz/2|=25$ (dots), 15 (dashes) and 5\gev\ (solid).
For $|\mchi-\mz/2|\gsim15\gev$ our calculation of $\abundchi$ is sufficiently
reliable. In window (b) the same for the light Higgs $h$.
}
\label{casetwodetails:fig}
\end{figure}

\begin{figure}[tbhp]
\caption{ Scaling behavior of the fine-tuning constant with the SUSY
scale. The
solid line represents a fine-tuning of 50 which typically corresponds to
$\msq, \mgluino\lsim1\tev$. The other lines are (left to right)
for 1, 10, 100, 500, and 1000. Here we have taken $\mt=145\gev$, $\tanb=5$,
$\azero/\mzero=-1$, and $\mu<0$.}
\label{ft:fig}
\end{figure}

\begin{figure}[tbhp]
\caption{ Scatter plot of (a) $\mgluino$, (b) $m_{\widetilde u_L}$, (c)
$\mzero$, and (d) $\mhalf$ vs.
fine-tuning for solutions consistent with all applied constraints.
Notice that the cut $f\leq50$ typically gives sparticle masses
$\msq,\mgluino\lsim1\tev$ but in some cases (all of which have large $\tanb$)
they can be significantly heavier.
}
\label{ftscatter:fig}
\end{figure}

\begin{figure}[tbhp]
\caption{ Scatter plot of $\bzero/\mzero$ vs.~$\azero/\mzero$ for all
allowed solutions (COMPASS) with $\mt=145\gev$. The quantized appearance is due
to
numerical sampling and is not significant.
}
\label{bzeroazero:fig}
\end{figure}

\begin{figure}[tbhp]
\caption{ Plots of the ($\mhalf,\mzero$) parameter space showing regions
excluded by lack of EWSB (labelled E), LSP not being the neutralino (L), and
the age of the Universe (A),
for (a) no 1-loop contributions to $\vhiggs$ and (b) leading 1-loop
contributions to $\vhiggs$. See text for discussion of full 1-loop
contributions to $\vhiggs$. For these plots we have taken $\mt=145\gev$,
$\tanb=5$, $\azero/\mzero=-1$ and $\sgnmu=-1$ (compare
Fig.~\protect\ref{casetwo:fig}a).}
\label{veffs:fig}
\end{figure}

\begin{figure}[tbhp]
\caption{ Lowest $\mhl$ versus
$\protect\sqrt{{\mstopone\mstoptwo}}$ for
all acceptable solutions with $\mt=145\gev$.
We allow $\mhl<60\gev$ for
the purposes of this graph only.
}
\label{apps:fig1}
\end{figure}

\begin{figure}[tbhp]
\caption{
Plot of $\mhl$ vs.~$\tanb$ for all acceptable solutions with
$\mt=145\gev$.  The solid vertical bands express the range in $\mhl$
for a given $\tanb$.
The dotted lines show the clear envelope
of $\mhl$ vs.~$\tanb$ that we obtain in the CMSSM.
Note that if $\mhl\lsim 85\gev$
then $\tanb\lsim 5$.
The discretization of $\tanb$ is merely
from numerical sampling and is not physically significant.}
\label{apps:fig2a}
\end{figure}

\begin{figure}[tbhp]
\caption{ Scatter plot of $\mhl$ vs.~$\protect\sqrt{\,|B\mu|}$ for
all acceptable solutions with $\mt=145\gev$ and $\tanb$=5.
Note that $\protect\sqrt{\,|B\mu|}$ is usually larger than $\mz$.
}
\label{apps:fig2b}
\end{figure}

\clearpage

\begin{figure}[tbhp]
\caption{ Histogram of $\mhl$ for all acceptable solutions
with
$\mt=145\gev$.}
\label{apps:fig11}
\end{figure}

\begin{figure}[tbhp]
\caption{ Percentage of acceptable solutions for $\mt=145\gev$
with $\hl$ being
detectable at LEP~II
versus center of mass energy (GeV).}
\label{apps:fig3}
\end{figure}

\begin{figure}[tbhp]
\caption{  Scatter plot of $\tanb$ vs.~$\mha$ for all
acceptable
solutions with $\mt$=145\gev.  The band structure is due to numerical
sampling and is not physically significant.  The dotted triangular
region is the approximate region in which it is difficult
to detect at least one Higgs~\protect\cite{kunszt7}.}
\label{apps:fig4}
\end{figure}

\begin{figure}[tbhp]
\caption{ Scatter plot of $\mhl$ vs.~$\mha$ for all
acceptable
solutions with $\mt$=145\gev.  The various bands correspond to different
choices of $\tanb$. The gap in the lower portion of the plot is due to
solutions
with $1.5<\tanb<3$ which were missed due to our finite grid.
}
\label{apps:mhma}
\end{figure}

\begin{figure}[tbhp]
\caption{ The ratio of the top Yukawa to the bottom Yukawa
couplings
at the GUT scale vs.~$\tanb$ for all acceptable solutions with
$\mt=145\gev$.  We see that
$\left( {h_{t0}\over h_{b0}}\right) \tanb \simeq 60$
is not a bad description, so
the running cannot help much to account for the large size of the $\mt/\mb$
ratio; it must
be imposed either via $\left( {h_{t0}\over h_{b0}}\right)$
or via $\tanb$.  If $\tanb$ is not large,
as we might expect, then a large $\left( {h_{t0}\over h_{b0}}\right)$
is required as input.}
\label{apps:fig5}
\end{figure}

\begin{figure}[tbhp]
\caption{ Histogram of BR($b\to s\gamma$) for all otherwise
acceptable solutions with $\mt=145\gev$.  The dotted lines indicate the
upper and lower bounds on BR($b\to s\gamma$) imposed as discussed
in the text. The SM prediction is $3.1\times10^{-4}$ corresponding to the
formulas of Ref.~\protect\cite{barbieri7}. More recent QCD correction
estimates~\protect\cite{adel7}\ lead to a SM
BR$(b\to s\gamma)\simeq4.3\times10^{-4}$.}
\label{apps:fig6}
\end{figure}

\begin{figure}[tbhp]
\caption{ Scatter plot of BR($b\to s\gamma$) vs.~$m_{H^\pm}$
for
all acceptable solutions with $\mt=145\gev$.  The faint banding visible
in the figure is from numerical sampling and is not physically significant.}
\label{apps:fig7}
\end{figure}

\begin{figure}[tbhp]
\caption{ BR($b\to s\gamma$) vs.~$\tanb$
for all acceptable solutions
with $\mt=145\gev$.  In order to demonstrate the density of points, $\tanb$
is slightly smeared around its numerically sampled value.}
\label{apps:fig8}
\end{figure}

\begin{figure}[tbhp]
\caption{ Regions of the
$(\mzero,\mhalf)$ plane to be explored
by SUSY searches at LEP~II if $\mt=145\gev$. Crosses (empty boxes)
mark the regions that will always (never) be accessible to LEP~II for any
combination of input
parameters. Filled boxes mark the regions accessible for some
combinations of parameters. Empty regions are excluded by our set of
constraints.
In window (a) we show the combination
of possible (direct) SUSY searches at LEP~II as described in the text. In
window (b) we show the LEP~II search potential for the chargino $\charone$
alone, and
in window (c) and (d) for the lightest Higgs assuming ability to find $\hl$
with
masses $\mhl<80\gev$ and
110\gev, respectively.
($\hl$ is a very SM-like Higgs.) Very large values of $\mhalf$
are disfavored by the fine-tuning constraint.
}
\label{apps:fig23}
\end{figure}

\begin{figure}[tbhp]
\caption{ $m_{\chi^{\pm}_1}-m_{\chi^0_1}$
vs.~$m_{\chi^{\pm}_1}$ for
all acceptable solutions with $\mt=145\gev$ and
$m_{\chi^{\pm}_1}<120\gev$.  The line represents
$m_{\chi^{\pm}_1}=2m_{\chi^0_1}$ which is approximately true for a gaugino-like
LSP. While the line is
an approximate description of the results it is not accurate enough
for detailed use.}
\label{apps:fig9}
\end{figure}

\begin{figure}[tbhp]
\caption{ Histogram of lepton or jet energy from
$\chi^{\pm}_1$
decay for acceptable solutions with $m_{\chi^{\pm}_1}<120\gev$.}
\label{apps:fig10}
\end{figure}

\begin{figure}[tbhp]
\caption{ Histogram of $R_b=\Gamma_{b\bar b}/\Gamma_{\rm
had}$
for all acceptable solutions with $\mt=145\gev$ and
$0.16<\abundchi <0.33$.  The central dotted line is the measured
value of $R_b$ and the outside dotted lines are the one standard
deviation errors on the measurement.  The dashed line is the
Standard Model calculated value of $R_b$ given $\mt=145\gev$.}
\label{apps:fig22}
\end{figure}

\begin{figure}[tbhp]
\caption{ Scatter plot in the plane ($\mu,M_2$) for
the acceptable solutions with $\mt=145\gev$. Notice a large concentration of
points below the diagonals
$M_2=|\mu|$  which shows that the LSP is gaugino-like
in most of the solutions.}
\label{mum2:fig}
\end{figure}

\begin{figure}[tbhp]
\caption{ Scatter plot of the LSP wavefunction for
the acceptable solutions for which
$0.16<\abundchi<0.33$ (MDM scenario), and for $\mt=145\gev$
and tan$\beta =10$.  Each solution contributes
scatter plot points corresponding to its $\widetilde{B}$,
$\widetilde{W}$,
$\widetilde{H}_d$, and $\widetilde{H}_u$ components.  Note that the LSP is
mainly $\widetilde{B}$ in all these solutions.}
\label{apps:fig20}
\end{figure}

\begin{figure}[tbhp]
\caption{ Histogram of $\abundchi$
for all acceptable solutions (with
an initial selecting cut $\abundchi<1$ imposed) for $\mt=145\gev$.
Notice a strong peak around $\abundchi\approx0.1$ showing that in
the CMSSM
the MDM
scenario is somewhat more favored relative to the CDM one.
}
\label{apps:fig21}
\end{figure}

\begin{figure}[tbhp]
\caption{ $\abundchi$ vs.~(a) $\mzero$, (b) $\mhalf$,
(c) $\tanb$, and (d) $\mu$, for otherwise
acceptable solutions with $\mt=145\gev$.
The bound $\abundchi<1$ comes from the age of the Universe of 10
billion years or more. The ranges $0.16\lsim\abundchi\lsim0.33$
and $0.25\lsim\abundchi\lsim0.5$ are favored by the MDM and CDM
scenarios, respectively.
The banding in $\mzero$, $\mhalf$ and $\tanb$ is
from numerical sampling and is not physically significant.}
\label{apps:relic:fig}
\end{figure}

\begin{figure}[tbhp]
\caption{ Scatter plot of mass vs.~particle type for all
acceptable
solutions in our data set with $\mt=145\gev$ and $f\leq50$.
The horizontal bands are due
to numerical sampling and are not of significance.}
\label{apps:fig13}
\end{figure}

\end{document}